\documentclass[msom,nonblindrev]{informs4} %
\OneAndAHalfSpacedXI %

\usepackage[table]{xcolor}
\usepackage{graphicx}
\usepackage{caption}
\usepackage[skip=0.5ex]{subcaption}
\usepackage{soul}
\usepackage{placeins}
\usepackage{amsmath}
\usepackage{afterpage}

\usepackage{changes}
\usepackage{tikz}
\usepackage{bm}

\usetikzlibrary{calc}
\usepackage{pgfplots}
\pgfplotsset{compat=1.15}
\usepackage{caption}
\usepackage{comment}
\usepackage{multirow}
\usepackage{bbm}

\usepackage{amsmath}
\usepackage{enumitem}
\usepackage{arydshln}
\usepackage{float}
\usepackage{subcaption}
\usepackage{booktabs}
\usepackage{framed}
\newenvironment{instructionbox}{%
  \begin{framed}\footnotesize\setlength{\parindent}{0pt}\setlength{\parskip}{0.3\baselineskip}\setlength{\baselineskip}{0.9\baselineskip}%
}{%
  \end{framed}%
}

\newenvironment{hyp}[1]{%
  \par\addvspace{6pt plus 3pt minus 2pt}%
  \noindent\hspace*{1em}{\normalfont\TheoremHeaderFont #1.}\hskip0.5em\itshape\ignorespaces}{%
  \par\addvspace{6pt plus 3pt minus 2pt}}
\newcommand{\hypref}[1]{{\normalfont\TheoremHeaderFont #1}}
\newsavebox{\tabnotebox}

\usepackage{natbib}
 \bibpunct[, ]{(}{)}{,}{a}{}{,}%
 \def\bibfont{\footnotesize\linespread{1.13}\selectfont}%
\counterwithin{equation}{section}

\newcolumntype{C}[1]{>{\centering\let\newline\\\arraybackslash\hspace{0pt}}m{#1}}
\newcolumntype{L}[1]{>{\raggedright\let\newline\\\arraybackslash\hspace{0pt}}m{#1}}
\newcolumntype{R}[1]{>{\raggedleft\let\newline\\\arraybackslash\hspace{0pt}}m{#1}}

\newcounter{marginparcounter}

\TheoremsNumberedThrough     %
\ECRepeatTheorems

\begin{document}

\RUNAUTHOR{Kagan}

\RUNTITLE{What's in a Queue?}

\RRHFirstLine{\bf\theRUNAUTHOR:\enskip {\it\theRUNTITLE}}
\LRHFirstLine{\bf\theRUNAUTHOR:\enskip {\it\theRUNTITLE}}
\def\setoddRH{\hbox to \textwidth{\fs.7.8.\tabcolsep0pt
  \begin{tabular*}{\textwidth}[b]{l@{\extracolsep\fill}r}
  {\theRRHFirstLine}& \raisebox{0pt}[0pt][0pt]{\fs.10.10.\thepage}\\[-4pt]
  \rlap{\VRHDW{0.5pt}{0pt}{\textwidth}}&\\
  \end{tabular*}}}
\def\setevenRH{\hbox to \textwidth{\fs.7.8.\tabcolsep0pt
  \begin{tabular*}{\textwidth}[b]{l@{\extracolsep\fill}r}
  \raisebox{0pt}[0pt][0pt]{\fs.10.10.\thepage}&{\theLRHFirstLine}\\[-4pt]
  \rlap{\VRHDW{0.5pt}{0pt}{\textwidth}}&\\
  \end{tabular*}}}

\def\HOOKtop{\vspace*{-0.95cm}}

\TITLE{What's in a Queue? An Experimental Study of Job Ordering, Autonomy and Queue Visibility}

\ARTICLEAUTHORS{%
\AUTHOR{Evgeny Kagan}
\AFF{Carey Business School, Johns Hopkins University, \EMAIL{ekagan@jhu.edu}}
} %

\ABSTRACT{%
\textbf{Problem Definition:} How a queue of jobs is arranged and presented to workers is an important design problem in service operations. This includes choosing the order in which jobs are performed, how much say workers have in setting that order, and how much queue and arrival information workers receive. \textbf{Methodology/Results:} To better understand how queue design (job ordering, autonomy, visibility) affects worker performance (speed, quality), we run a series of pre-registered online experiments. We use a new, real-effort task in which workers fulfill order-picking jobs of varying complexity that arrive dynamically over time. Our results are as follows: (1) When workers choose their own picking order, we reproduce the field finding that Easy First (EF) ordering is associated with worse performance than First-in-first-out (FIFO), and show that this is mainly due to worker self-selection rather than due to the ordering itself; (2) Exogenously imposed EF ordering improves work quality (picking accuracy) relative to both FIFO and discretionary ordering; (3) Imposing an ordering may reduce speed for the most capable workers; (4) Seeing a new job arrival leads to a short-term productivity burst; however, removing job arrival and queue information altogether does not affect performance in the long term. \textbf{Managerial implications:} Our results provide guidance on which queue design works best for a given performance goal (speed or quality) and worker ability level. We also identify personality measures that can help managers screen for error-prone workers.}


\maketitle

\vspace{0.11cm}

\section{Introduction}\label{sec:intro}

How a queue of jobs is arranged and presented to workers is a key design choice in service operations. In this paper we seek to inform this design choice by addressing three questions that are of interest to both researchers and practitioners. \textit{In what order} should jobs be performed? \textit{Who} should make this decision? And \textit{how much visibility} should workers have into the queue and the job arrival process? Practice varies on all three questions. Often, jobs are completed in the order of arrival: First-in-first-out (FIFO) is the classic service-queue discipline \citep{gross1998} and is the dominant design across many service settings, from walk-in counters \citep{larson1987} to call centers \citep{gans2003}. In other settings, easier or quicker jobs are completed first, as in the shortest-processing-time or Easy First (EF) rule often used to sequence jobs in manufacturing \citep{conway1967}. Workers may also have varying degrees of discretion over job sequencing. For example, some warehouses use pick lists, where a worker chooses the route and sequence in which to collect items, whereas others use voice-directed picking, which removes this discretion \citep{netsuite2026voicepicking}. Finally, service providers differ in terms of how much of the queue the worker can see. For example, some call centers share with their agents a real-time dashboard showing when new callers join the queue and how many are currently waiting \citep{kent2026wallboard}, while others simply route the next caller to the appropriate agent \citep{genesys2026acd}.  

Task ordering has received some attention in the operations literature, mainly in healthcare settings. The default rule is often FIFO, but when given discretion, workers frequently deviate from it and pick easier jobs. \citet{kc2020} document this behavior in a hospital setting and show that it is more prevalent under high workload and is associated with lower complexity-adjusted throughput. In the same vein, \citet{ibanez2018} show that radiologists who deviate from FIFO are slower than those who follow the prescribed order. A common reading of these results is that  worker discretion should be limited. Even less is known about the effects of queue visibility on worker performance. The existing evidence is largely based on multi-server settings with a shared queue, where workers may have an incentive to free ride and queue length can serve as a signal of other servers' effort \citep{shunko2018,rosokha2024}. 

Experiments can complement field studies of queue-based work in several ways. First, worker performance in most settings has two relevant dimensions: speed and quality \citep{hopp2007}. Speed (or throughput) measures are often readily available in  field data sets, while quality measures are more difficult to obtain. A controlled experiment can add an objective quality measure. Second, job sequencing rules are often based on a common organizational policy and do not vary within or across workers; as a result, there is little exogenous variation to exploit. When workers do have discretion, sequence and performance are often endogenously determined by worker type and can be affected by situational factors (such as congestion, which itself can be endogenous). Experiments can help establish causality by imposing an exogenous sequence and by exogenously controlling arrival rates. Third, field data  rarely include detailed worker characteristics (e.g., baseline ability, preferences or personality traits), which makes it difficult to characterize heterogeneous effects of different queue designs. In an experiment, personality traits and baseline ability measures can be collected prior to treatment assignment, and treatment effects can be evaluated for different types of workers. To our knowledge, no prior work has attempted to experimentally address these issues.%

\subsection{Experiment Design and Preview of Results}
To examine how queue design affects performance, we develop a new experimental task that captures key features of service work. Participants individually complete a fixed set of order-picking jobs of varying complexity (Small, Medium, and Large) that arrive dynamically over time. The task is representative of a range of settings including customer support, IT helpdesks, warehouse picking, clerical processing, and other ticket-based work and has the following features:
\begin{itemize}[nosep]
    \item Prior to treatment assignment, all workers complete the same set of jobs in the same, pre-determined order. This gives us a baseline measure of worker ability, which provides an important control and allows us to examine heterogeneous treatment effects.
    \item Jobs arrive dynamically over the course of a round, creating a realistic queue environment and providing within-worker variation in queue length, arrival intensity, and queue composition. The job arrival schedule itself is fixed and identical across workers and conditions.  
    \item In some conditions, workers choose which job to work on next; in others, the sequence is imposed by the experimenter. This allows us to separate causal effects from selection.
    \item The task requires real effort. Furthermore, some rounds require the worker to submit jobs perfectly (i.e., the worker needs to redo the job if there are errors), while other rounds allow errors. This produces a separate measure for speed (with quality being fixed) and one for quality.
    \item Each worker processes their own, independent queue and must complete all jobs that arrive. This avoids social and strategic considerations that can arise when multiple servers share a workload or when workers can skip jobs or delegate to others, and allows us to focus on the effects of different queue designs on individual productivity.
\end{itemize}

We use this experimental approach to conduct a series of pre-registered experiments on the Prolific platform. In the first two experiments we focus on discretionary ordering. In particular, in Experiment 1a workers are free to grab jobs from their queue as they see fit. We will refer to this treatment as \textit{Endog} because the job sequences are endogenously determined by the worker. Experiment 1b is similar, but workers must commit to a sequence (EF or FIFO) before starting to work on their queue. We will refer to this treatment as \textit{Pre-commit}. In the remaining experiments (2a and 2b) we examine exogenously imposed job sequences. In Experiment 2a we introduce the exogenous EF (\textit{Exog-EF}) and exogenous FIFO (\textit{Exog-FIFO}) treatments. In Experiment 2b we remove queue visibility. Please see Table~\ref{tab:conditions} for a tabular view of our treatments.
\begin{table}[b]
\renewcommand{\arraystretch}{1.1}
\centering
\caption{Summary of Experimental Conditions}
\label{tab:conditions}
\footnotesize
\begin{tabular}{llcccc}
\toprule
 & Condition & Ordering & Autonomy & Visibility \& arrival alerts & $N$ \\
\midrule
\multicolumn{6}{l}{\textbf{Experiment 1: Worker Chooses Job Sequence}} \\
Exp.~1a & \textit{Endog} & Worker's choice & Full (real-time) & Full queue + arrival alerts & 100 \\
Exp.~1b & \textit{Pre-commit} & Worker's choice & Per-round choice & Full queue + arrival alerts & 111 \\
\addlinespace
\multicolumn{6}{l}{\textbf{Experiment 2: Role of Autonomy and Queue Visibility}} \\
Exp.~2a & \textit{Endog} (Replication of Exp. 1a) & -- & -- & -- & 101 \\
Exp.~2a & \textit{Exog-EF} & Easy First (EF) & None & Full queue + arrival alerts & 101 \\
Exp.~2a & \textit{Exog-FIFO} & First-in-first-out (FIFO) & None & Full queue + arrival alerts & 107 \\
Exp.~2b & \textit{Invisible-FIFO} & First-in-first-out (FIFO)  & None & Current job only, no arrival alerts & 100 \\
\bottomrule
\end{tabular}
\end{table}

Experiment~1 yields three sets of results. First, we are able to replicate several key findings from prior field and lab studies. In particular, consistent with \cite{edie1954,schultz1998,kc2009,ibanez2018} and \cite{kc2020}, we show that worker behavior is load-adaptive: as workload increases workers speed up and cherry-pick more. In particular, following a new job arrival, workers are more than twice as likely to cherry-pick easy jobs and complete jobs about 20\% faster (adjusted for complexity). Second, easier jobs that were picked out of sequence produce higher error rates. However, the relationship between cherry-picking and errors is not causal in our data: after we include worker fixed effects, cherry-picking is no longer associated with more errors. That is, rather than the act of cherry-picking causing errors, cherry-picking and high error rates are both markers of a common worker type. Third, in Experiment~1b we show that the link between choosing EF and making more errors persists when job sequence is selected by the worker before work begins (i.e., at the beginning of a round). This suggests that the inclination to work easy-first (and the associated decline in work quality) is not just an impulsive reaction to momentary stress or workload, but a deliberate preference.

Experiment~1 results raise three follow-up questions. First, can we characterize the worker types that choose easy-first sequences and submit jobs of lower quality? Second, can a firm improve performance by \textit{requiring} workers to follow a particular sequence rather than letting them choose one? Third, given that recent arrivals produce short-term productivity bursts, does removing queue and arrival information help or hurt performance over the whole work period? Experiment 2 addresses these questions. To answer the first question, we re-run the \textit{Endog} treatment and add a battery of personality measures administered before the main task. We find that most personality measures are weak predictors of behavior, with one clear exception: risk-loving workers prefer EF and submit more error-prone work. For the second question, we exogenously impose the job sequence (FIFO or EF, depending on treatment) rather than letting workers choose it. We find that imposing EF reduces the error rate relative to both \textit{Endog} and \textit{Exog-FIFO} by about 30\%. This improvement is mostly concentrated among the weaker performers who appear to benefit from a gradual increase in job complexity. Furthermore, imposing a fixed sequence (particularly FIFO) does not affect speed on average, but it slows down the ablest workers. Finally, we vary queue visibility while holding job order and autonomy fixed and show that both speed and quality (error rates) are essentially unchanged relative to the visible condition. This suggests that the speedup in response to high workload is a momentary effect that is offset by slowdowns during quieter stretches.

\subsection{Contributions}
We make three contributions. First, we develop a new experimental approach for studying the effects of different queue designs on individual productivity. Our approach allows us to observe both speed and quality, adding a richer measure of performance relative to field studies of discretionary ordering, which tend to focus on throughput or complexity-adjusted productivity \citep[e.g.,][]{ibanez2018, kc2020}, and relative to existing lab studies, which focus mostly on speed \citep[e.g.,][]{schultz1999,shunko2018}. %

Second, our experiment design allows us to identify a new mechanism driving the previously documented relationship between easy-first ordering and weaker performance \citep{ibanez2018, kc2020}. One of the mechanisms proposed in that literature is that the worker must repeatedly evaluate the queue and reassess job complexity before deciding what to do next, and this can lower productivity by detracting from their time and attention. In our setting, job complexity is clearly displayed to workers, so the cost of inspecting the queue is minimal. Nevertheless, we also find that easy-first ordering is associated with poor performance. We identify a new, selection-based mechanism behind that performance gap: a specific worker type that tends to grab easy jobs first and submits low-quality work. Furthermore, exogenously imposing easy-first ordering does not harm performance and may even help reduce error rates, which has important managerial implications. Our results thus add to the field findings by identifying a new mechanism and the conditions under which easy-first ordering helps vs. hurts performance.

Third, we add to the behavioral queueing literature studying the effects of queue visibility on performance. Whereas prior queueing experiments show that queue design can affect effort through strategic behaviors such as social loafing \citep{shunko2018, rosokha2024} or through blocking/starving the assembly line \citep{schultz1999}, we study a single worker completing their own, individual queue and find that hiding the queue  in this setting leaves performance essentially unchanged.

\section{Related Literature}\label{sec:literature}
Our work builds on and contributes to the literature on worker behaviors under different workloads (\textsection 2.1), and the three design choices manipulated in our experiments: the order in which jobs are done (\textsection 2.2), who controls that order (\textsection 2.3), and how much of the queue workers can see (\textsection 2.4).

\subsection{Workload Effects}
Classic queueing models treat service rates as fixed and exogenously given \citep{kleinrock1976, gross1998}. A growing literature \citep[][]{george2001,dong2015, allon2018,ingolfsson2023} questions this assumption and suggests that server behavior is load-adaptive. Within that, a substantial body of work supports the idea of state-dependent service rates \citep[see][for a review]{delasay2019}. The most common finding is that workers process jobs faster when the system is busier \citep{edie1954, schultz1998, kc2009, tan2012}, although under very high loads workers may become fatigued and slow down again \citep{delasay2016,berry2017}. Workers also delegate more under higher load: when given the opportunity, they transfer more requests to others \citep{hathaway2023} and rely more on algorithmic advice \citep{snyder2025}. The bulk of the existing evidence is focused on speed or throughput and does not consider potential trade-offs between speed and quality. Nonetheless, this literature yields two testable predictions for our experiments. First, higher workload salience should improve performance (speed). Second, higher workload salience should increase the tendency to pick easier jobs.

\subsection{Job Sequencing}
Scheduling theory suggests that easy-first sequences (often referred to as shortest expected processing time in scheduling) minimize average flow time \citep{schrage1966}. The behavioral literature furthermore  suggests that easy-first has a broad appeal to workers in a variety of settings and is sometimes linked to worse performance.  \cite{kc2020} show that completing easy jobs first is common and lowers complexity-adjusted throughput in a hospital setting. \cite{ibanez2018} find that deviating from the assigned order reduces radiologists' throughput because inspecting the queue to choose the next case takes time. A common reading of these results is that workers' discretion over ordering should be limited (see \textsection 2.3 for more on worker autonomy).\footnote{In some settings starting with the easier job has been shown to improve performance, particularly in innovation-related settings where workers need time to get familiarized with a new set of materials or constraints \citep{kagan2025}. In our experiments, workers complete 52 jobs with identical structure over the course of six rounds, so that such learning effects should not play a big role. Relatedly, \citet{siemsen2008} and \cite{katok2011} show that workers may select more difficult jobs to signal their type and build a reputation. We abstract away from reputational concerns and other strategic motivations and focus on individual worker productivity.} Also related is \citet{rusou2020}, who give participants a fixed set of data-retrieval jobs of varying size to complete within a time limit, and find that people tend to choose smaller, less effortful jobs even when larger ones pay more per unit of time. A related small-task bias is documented in \cite{pape2024} in the project management setting. Different from these papers, we study task selection in a service setting with a dynamic queue and examine effects on productivity in a real-effort task, rather than strategic decisions. Furthermore, unlike much of prior work, we measure both quality and speed. Nonetheless, this literature motivates our hypothesis that easy-first ordering hurts performance.

\subsection{Worker Autonomy}
The literature discussed in \textsection 2.2 studies what workers do when they order their own work; far less is known about whether giving them that discretion helps or hurts performance. \citet{hopp2007} develop a model where workers can decide for themselves when a job is finished and may lower quality to manage their workload. We will also examine speed-quality trade-offs in our data. Empirically, imposing structure has been shown to improve outcomes in two different settings. \citet{bray2016} show that when judges stopped juggling multiple cases at once and instead were required to finish the oldest case before opening a new one, average case durations and appeal rates fell. \cite{kagan2018} find that leaving time allocation decisions to workers can add a cognitive burden and that designers who choose when to move from ideation to execution perform worse than those forced to follow an external schedule. Importantly, none of this work varies who controls the \textit{sequence} of a fixed set of jobs as we do in the current paper.

\subsection{Queue Visibility}
The third design choice is \textit{what} workers can see about their queue. Most related evidence comes from multi-server settings, where several servers draw from a shared queue or work side by side, and queue visibility can provide a signal into other servers' effort. \citet{shunko2018} vary both the queue structure (parallel vs.\ pooled) and queue-length visibility, and show that servers who can see the queue length work faster than those for whom it is hidden. \citet{rosokha2024} show that queue visibility is a key driver of equilibrium strategies that sustain effort when servers share a queue and are able to free ride. Visibility may also matter for an individual server through the perception of their own workload. \citet{songARRoels2024} find that dedicated queues improve speed relative to pooled queues because a dedicated queue makes servers feel more ownership over the work. However, in their design, only queue configuration is varied, while visibility is held fixed. Experiment 2b tests the visibility effect in the context of \textit{individual} productivity rather than a shared queue setting, so free riding and cooperation motives cannot affect behavior. If the speed-up under load documented in \textsection 2.1 requires seeing the load, workers who cannot see the queue should slow down. If instead the visible backlog distracts from the job at hand, hiding it could improve performance. %

\section{Experiment 1: Discretionary Job Ordering}\label{sec:exp1a}
Experiment~1 consists of two parts: Experiment 1a and Experiment 1b. In both of these experiments participants are given the discretion to choose their own job sequences, with the difference that  in Experiment~1a they pick jobs from the list one job at a time as they work, whereas in Experiment~1b they commit to a sequence (Easy First or FIFO) before each round begins.

\subsection{Experiment 1a}
Experiment 1a was conducted on Prolific in January 2026.\footnote{Please see \S\ref{ec:prereg} for pre-registration and \S\ref{ec:pilot} for details on the pilots conducted prior to the main experiment.} The experiment consisted of a single treatment, which we will refer to as \textit{Endog} (Endogenous, i.e., discretionary task ordering). Average time spent in the experiment was 30.2 minutes (instructions and exit survey: 8.5 minutes; main task: 21.7 minutes). Sample sizes, exclusions, and demographic composition for all conditions reported in the paper are summarized in Table~\ref{tab:exclusions} in the e-companion.

\subsubsection{Experimental Task}
We use a real-effort task designed to reproduce the key features of ticket or job-based work where the worker has the discretion to choose the sequence in which the jobs are performed, as is common in many jobs in retail, IT work, warehousing or certain healthcare settings (e.g., radiology). Participants complete a series of ``jobs''.  Each job is a virtual grocery order and requires selecting correct items (and correct quantity) from a catalog. Items include basic grocery and hygiene items such as milk, ketchup, toothpaste, etc. To facilitate learning and limit order picking time, the catalog includes a total of 29 items. There are three types of jobs of varying complexity:
\begin{itemize}[nosep]
    \item \textbf{Small}: 1 item total 
    \item \textbf{Medium}: 3 items total (2 distinct items)  
    \item \textbf{Large}: 10 items total (3 or 4 distinct items)  
\end{itemize}

Job size, difficulty, and expected duration are generally higher for Medium jobs than for Small jobs and are higher for Large jobs than for Medium jobs. We therefore use ``easier'' and ``smaller'' interchangeably and refer to sequences that prioritize easier jobs as Easy First (EF).

\begin{figure}[b!]
\centering
\caption{Experimental Interface}
\begin{subfigure}[b]{0.8\textwidth}
\centering
\caption{Job List}
\label{fig:screenshots}
\includegraphics[width=\textwidth]{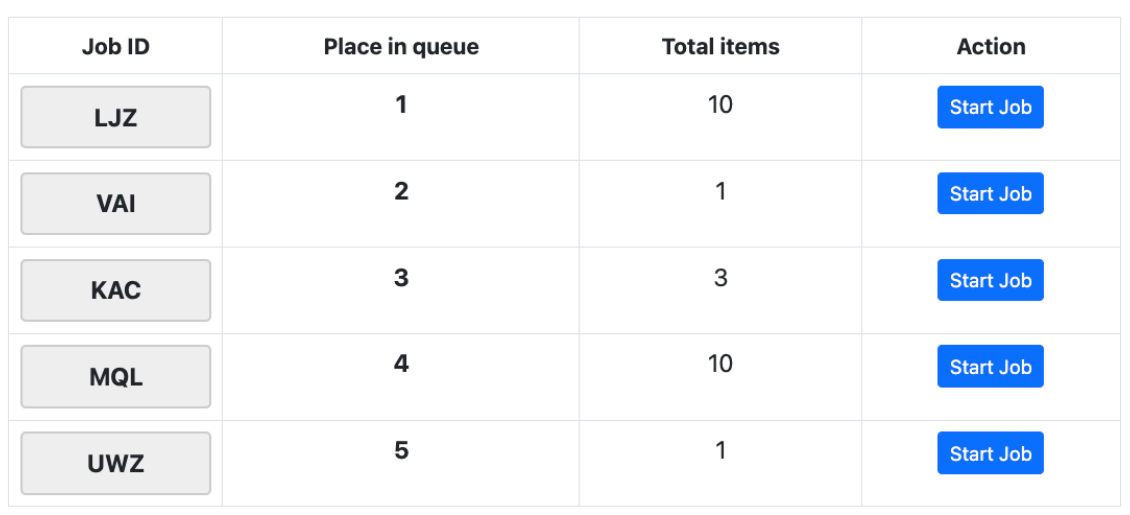}
\end{subfigure}

\vspace{1em}

\begin{subfigure}[b]{0.8\textwidth}
\centering
\caption{Sample Job (Abbreviated)}
\label{fig:samplejob}
\includegraphics[width=\textwidth]{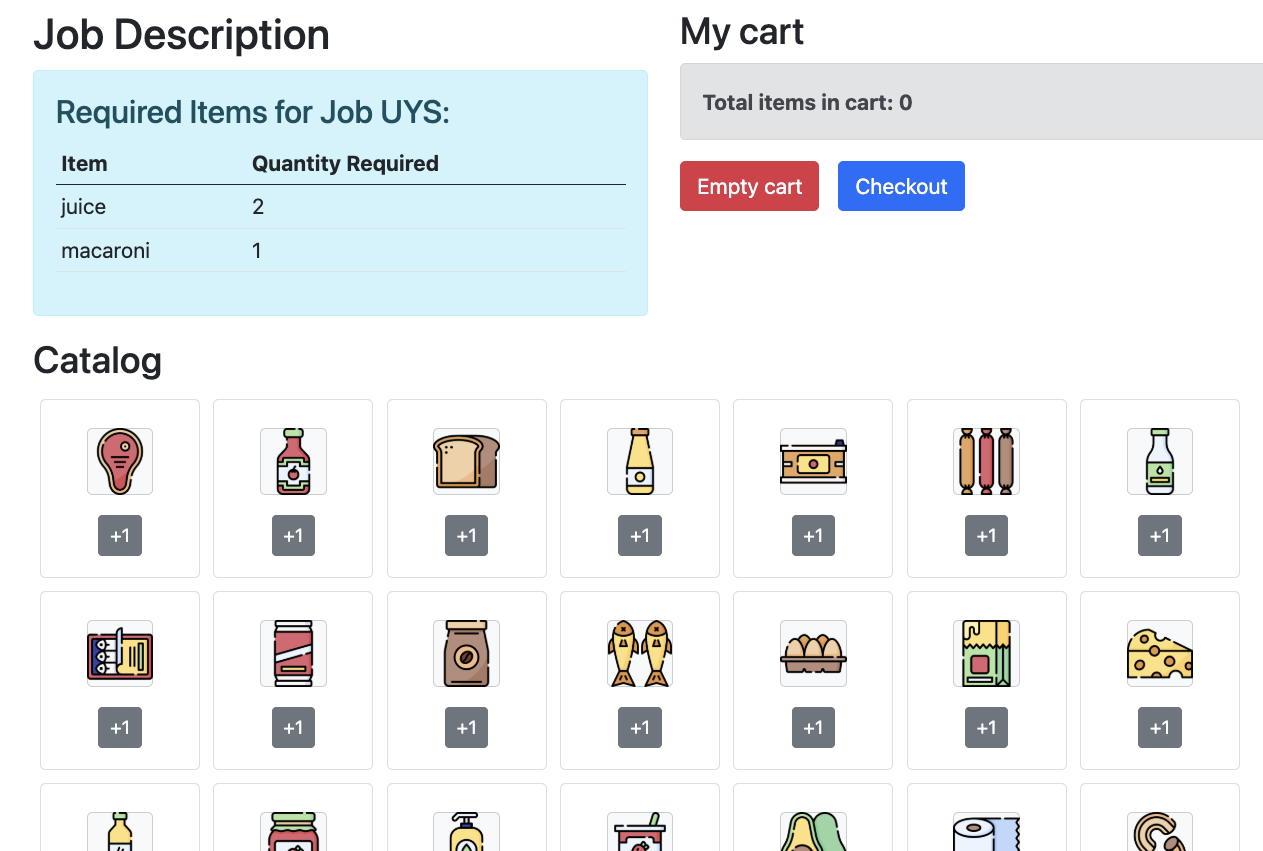}
\end{subfigure}
\end{figure}

All participants complete 42 jobs in the four main rounds, of which 14 are Small, 14 are Medium, and 14 are Large. Jobs arrive dynamically according to a pre-programmed arrival schedule and, upon arrival, each job is announced to the worker via a banner at the top of the page. The arrival schedule is fixed across workers and does not depend on their behavior or performance. See Figure~\ref{fig:screenshots} for the job selection screen. After a job has arrived, it is added at the bottom of the queue (in last place). Workers cannot skip jobs but they can select the order in which they complete the jobs. Arrival rates were set such that workers of all abilities have ample cherry-picking opportunities throughout the work horizon (arrival times were calibrated based on pilot results; see \S\ref{ec:pilot}).\footnote{The calibration worked as intended: $100\%$ of workers had at least one job in the queue at every selection, $89\%$ of selections had two or more jobs, and $80\%$ had a queue of mixed difficulties, i.e., a cherry-picking opportunity was available. Additionally, we used two different randomly drawn arrival sequences for each main round. This ensures that the results are not driven by any particular item appearing at a particular point in the experiment (Also see the job-content robustness checks in \S\ref{ec:rob:controls}).} After clicking on the ``Start Job'' button, workers are redirected to the order picking screen for the chosen job. Products can be added to the cart by clicking on the ``+1'' button under the product image. Figure~\ref{fig:samplejob} shows a screenshot of a medium-size job. Product names were intentionally hidden to increase difficulty (had they been displayed, the task would have been near-trivial). Product positions were randomized across jobs to increase difficulty.

\subsubsection{Protocol, Manipulations and Compensation}
\begin{figure}[b!]
\centering
\caption{Experimental Protocol}
\includegraphics[width=\textwidth]{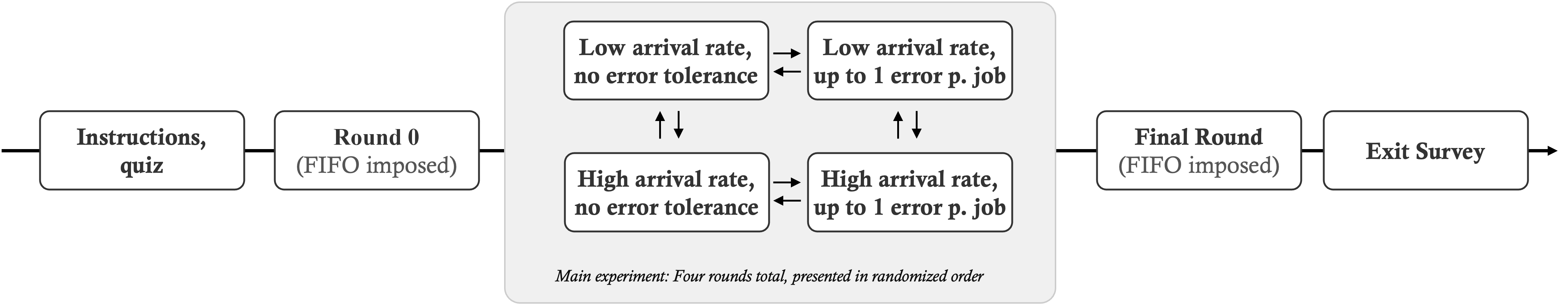}
\label{fig:design}
\end{figure}

Figure~\ref{fig:design} summarizes the experimental protocol. After providing informed consent, participants went through instructions and comprehension screening. The experiment itself consisted of six rounds: Round 0 (Baseline), four main rounds, and a final round. In Round 0, five jobs were all available from the start and the completion order was fixed. Similarly, the final round also presented participants with five jobs, which had to be completed in a pre-determined order. These rounds allow us to assess baseline performance, as well as potential learning effects. Furthermore, there were a total of four main rounds, in which we manipulated two within-subject factors:
\begin{enumerate}[nosep]
    \item \textbf{Error tolerance}
        (zero-tolerance rounds vs.\ error-tolerant rounds):
        In zero-tolerance rounds, any error requires
        the participant to redo the job.
        In error-tolerant rounds, up to one error 
        per Medium or Large job is accepted
        without requiring a redo. Small jobs must always be completed without errors regardless of round type. 
        Half of participants experience
        zero-tolerance rounds first (Rounds 1--2)
        and error-tolerant rounds later (Rounds 3--4);
        the other half experience the reverse order.
    \item \textbf{Arrival rate} (low vs.\ high):
        In low-arrival-rate rounds, 9 jobs arrive over the course of the first 80 seconds;
        in high-arrival-rate rounds, 12 jobs arrive during the same period.  We use this within-subject manipulation to examine the effects of arrival rate. The order of rounds (low arrival rate first/high arrival rate first) was randomized within each block.
\end{enumerate}

Subjects were paid 5 cents for each completed job, where ``completed'' means that subjects had to redo a job until it was error-free in zero-tolerance rounds or had up to one error (missing item, wrong item, insufficient quantity) in error-tolerant rounds. Subjects also received a show-up fee of \$4. Furthermore, there was no time limit on completing the jobs, and subjects could not leave the experiment and get paid prior to completing all jobs. 

\subsubsection{Recruitment and Exclusions}\label{sec:exp1a:recruit}
The pre-registered target sample size was 100 subjects. Prolific was instructed to recruit US-based workers with a 99\%+ approval rate. Full recruitment criteria are reported in \S\ref{ec:prereg}. The pre-registered exclusion criteria were (i)~making four or more mistakes on the comprehension quiz and (ii)~self-reported technical issues that prevented normal completion of the task. After applying all screens, a total of 100 subjects entered the final dataset.\footnote{In addition to these pre-registered exclusions we apply one additional exclusion that we did not anticipate during pre-registration: we drop inattentive subjects who appear to have stepped away from the task for an extended period of time. Specifically, we flag a subject as inattentive if any of their main-round job selections occurred more than 60~seconds after the previous job was completed (an unusually long idle gap between finishing one job and starting the next). This filter excludes 3 subjects (with maximum between-job gaps of 60.2, 63.6, and 127.2~seconds), leaving $n=100$ in our analysis sample. The full-sample results (without the post-hoc attentiveness filter) are reported in \S\ref{ec:rob:full} and are similar in terms of effect sizes and significance levels.}

\subsubsection{Measurements and Hypotheses}
We focus on two types of outcomes. The first one is task performance, measured as (i) job completion time and (ii) error rate. Since one can work faster by making more errors in the error-tolerant rounds, we do not generally consider these rounds when measuring speed;  to get a clean measure of speed we instead look at completion times in the zero-tolerance rounds (where quality is held constant). Conversely, to measure work quality (error rates) we will focus on the error-tolerant rounds. In these rounds quality is more discretionary and participants face a trade-off between completing a job quickly but potentially making an error (since up to one error is allowed on Medium and Large jobs) vs.\ working more thoroughly to ensure that there are no errors. 

To develop hypotheses for how the state of the service system affects worker behavior, and for the relationships between task ordering and performance, we draw on two streams of research: the literature on workload and service speed, and the literature on task selection and performance, both of which were discussed in detail in \textsection 2. To recap, when workload increases, workers typically speed up \citep{edie1954, schultz1998, schultz1999, kc2009, tan2012, delasay2019} and pick easier jobs \citep{ibanez2018, kc2020}. We expect to see similar effects in our setting. Since we observe the time elapsed between each job arrival and the subsequent job selection and job performance, we can measure how workers respond to perceived arrival rate at a granular level. Additionally, arrival rate is exogenously varied across rounds (9 jobs vs.\ 12 jobs), providing further variation in workload.\footnote{Since we observe both queue length and the timing of recent arrivals, we can also disentangle the effects of queue buildup from perceived arrival rate; to do this, we will control for queue length in our regression analyses.} Formally:
 
\begin{hyp}{H1a (Workload and Performance)}
Higher workload salience (perceived arrival rate) leads to better performance (faster completion times and fewer errors).
\end{hyp}

\begin{hyp}{H1b (Workload and Job Ordering)}
Higher workload salience (perceived arrival rate) leads to more cherry-picking.
\end{hyp}

Our second measure of interest is related to job sequencing behavior. In particular, we use ``cherry-picking'' and ``easy-first sequencing'' interchangeably to describe picking a job that is easier (smaller) than the job at the front of the queue, i.e., the oldest waiting job. Importantly, workers must eventually complete every job; cherry-picking changes only the order in which jobs are completed.

On one hand, completing easy jobs quickly may help build momentum. Completing many small jobs first creates visible progress that could motivate continued effort. On the other hand, deferring difficult jobs leads to these jobs languishing in the queue, which may decrease motivation. Prior empirical work suggests negative consequences of cherry-picking easy jobs on performance \citep{ibanez2018, kc2020}. Following this literature we hypothesize that cherry-picking is associated with poor performance. However, we hypothesize that this could be due to two mechanisms: a \emph{selection} mechanism wherein workers who cherry-pick may be systematically different (e.g., lower ability, more easily distracted) from workers who follow FIFO, and a \emph{causal} mechanism wherein the act of deferring difficult jobs itself reduces performance, for example by allowing fatigue to accumulate, or by spending time browsing the queue. Therefore, we hypothesize:

\begin{hyp}{H2a (Job Ordering and Performance, Correlational)}
There is a negative correlation between cherry-picking and performance.
\end{hyp}

\begin{hyp}{H2b (Job Ordering and Performance, Causal)}
The negative relationship between cherry-picking and performance persists, once stable worker characteristics are absorbed by subject fixed effects.
\end{hyp}

\subsubsection{Results: Summary Statistics}
\begin{figure}[bt]
\centering
\caption{Experiment 1a: Summary Statistics}
\includegraphics[width=0.7\textwidth]{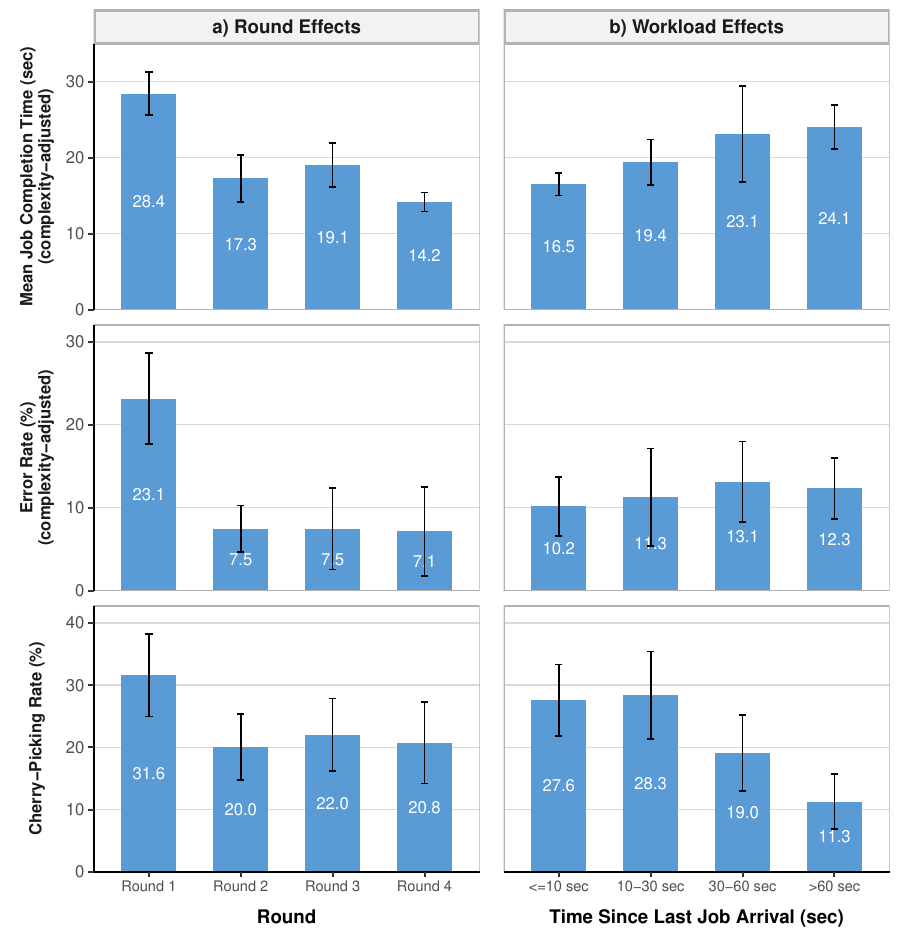}
\begin{minipage}{0.69\textwidth}\scriptsize
\textit{Notes:} Completion times and error rates are complexity-adjusted: within each bin, job-type means are weighted by the overall job-type shares. Error bars are 95\% CIs, clustered at the subject level.
\end{minipage}\label{fig:summary}
\end{figure}

We begin by examining broad patterns in the data across our three main outcomes: job completion times, error rates, and cherry-picking rates. Figure~\ref{fig:summary} presents these outcomes as functions of round (column a) and time since last job arrival (column b). Mean completion times decline from 28.4 seconds in Round~1 to 14.2 seconds in Round~4 (rank sum test, $p<0.001$), error rates drop sharply from 23.1\% to 7.1\% ($p<0.001$), and cherry-picking rates decrease from 31.6\% to 20.8\% ($p<0.001$). These patterns suggest that participants become both faster and more accurate with experience, while also becoming somewhat less prone to deviating from FIFO. The second column examines workload effects. We measure perceived workload as the time since the last job arrival: the more recent the last arrival, the more salient the workload. Completion times show a positive association with time since arrival, increasing from 16.5 seconds to 24.1 seconds ($p<0.001$), while cherry-picking rates fall from about $28\%$ following arrivals within the last 30 seconds to $11.3\%$ once more than 60 seconds have passed ($p<0.001$). Error rates, however, do not vary significantly with this workload measure ($p=0.59$), ranging between 10.2\% and 13.1\% without a clear trend. Overall, these comparisons provide some initial support for \hypref{H1}.

\begin{figure}[tb]
\centering
\caption{Job Completion Time and Error Rate by Cherry-Picking Status}
\includegraphics[width=0.65\textwidth]{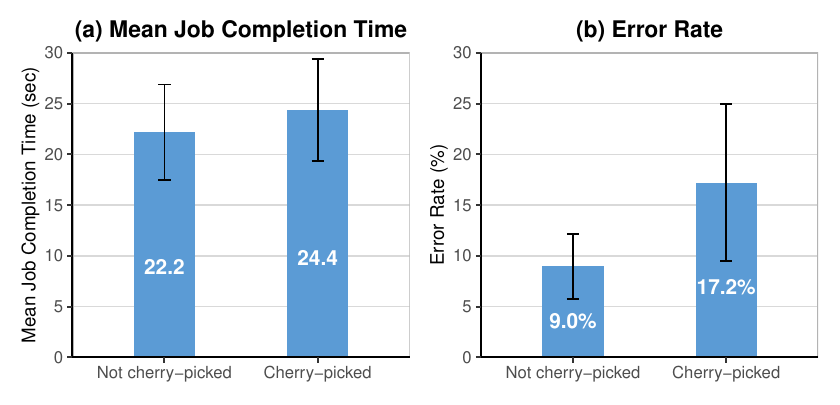}
\begin{minipage}{0.97\textwidth}\scriptsize
\textit{Notes:} Medium jobs only, restricted to selections made when a cherry-picking opportunity existed. ``Cherry-picked'' = the worker chose the job ahead of an older, larger job; ``Not cherry-picked'' = job was performed in FIFO order. Error bars are 95\% CIs, clustered at the subject level.
\end{minipage}
\label{fig:cp_task}
\end{figure}

Next consider the relationship between cherry-picking behavior and performance. We focus on Medium jobs because they can be cherry-picked (i.e., completed ahead of Large jobs) and also allow for errors (recall that Small jobs had to be completed without errors, see \textsection 3.1.2). Note that Large jobs cannot be cherry-picked (by definition), although we will later examine whether completing larger jobs ahead of easier ones is related to performance. Figure~\ref{fig:cp_task} compares jobs that were cherry-picked (selected when a larger job was waiting ahead of it in the queue) to jobs that were not. Panel~(a) shows that cherry-picked jobs take slightly longer to complete (24.4 vs.\ 22.2 seconds; rank sum test, $p=0.06$). Panel~(b) shows that cherry-picked  jobs have a much higher error rate (17.2\% vs.\ 9.0\%; rank sum test, $p=0.004$). That is, cherry-picking is associated with more errors. These comparisons provide some descriptive evidence in support of \hypref{H2a}, but they do not account for worker-level heterogeneity: workers who cherry-pick more may be systematically different from those who do not. We address this in the regression analysis in \textsection 3.1.6.

\subsubsection{Hypothesis Tests}\label{sec:exp1a:tests}
We begin by testing \hypref{H1}, i.e., the effect of workload salience on job completion time, error rates, and cherry-picking. Table~\ref{tab:reg:H1} reports the results. All specifications include round fixed effects and control for baseline ability, with standard errors clustered at the subject level; columns~(1)--(4) additionally include job-type fixed effects, which are not defined for the selection-level observations in columns~(5)--(6). Columns~(1)--(2) estimate OLS regressions of completion time; columns~(3)--(4) and (5)--(6) report log-odds coefficients from logit regressions with the error rate and cherry-picking being the dependent variables, respectively.\footnote{In Table~\ref{tab:reg:H1} we control for observed worker heterogeneity through baseline ability (Round~0 completion time and redo count). For robustness, Table~\ref{tab:reg:H1:fe} in \S\ref{ec:robustness} re-estimates the cherry-picking columns with subject fixed effects, using both a FE logit and a linear probability model. For further robustness, Table~\ref{tab:reg:H1:sku} replaces the job-type controls with fixed effect indicators for each of the 29 items in the catalog, so that the content of each job is controlled for. The results are similar in size and significance.} First, consider column (1): consistent with \hypref{H1a}, workers complete jobs faster following recent arrivals: the log specification shows that a one-unit increase in log time since the last job arrival increases completion time by 1.370 seconds ($p<0.01$), and the categorical specification in column~(2) shows that workers who experienced an arrival within the last 10 seconds complete jobs 4.714 seconds faster than those whose last arrival was more than 60 seconds ago ($p<0.01$).\footnote{In columns (1), (3) and (5), we use the log transformation because time since last arrival is heavily right-skewed (median 13 seconds, mean 49 seconds), so a linear specification puts a lot of weight on the long right tail of arrivals.  Nonetheless, we redo the analysis using a linear specification and find qualitatively similar results (results omitted to conserve space).} The speed gains are observed for the two most recent bins, which are indistinguishable from each other ($\leq$10 vs.\ 10--30 seconds, $p=0.47$). However, the speed gains are no longer observed once no new job has arrived for 30 seconds: workers whose last arrival was 30--60 seconds ago complete jobs just 1.643 seconds faster than the $>$60 second group, a difference that is not statistically significant ($p=0.16$). Note that these effects are adjusted for job type (Small vs. Medium vs. Large), which is included in the controls.  Columns~(3)--(4) show no relationship between arrival recency and error occurrence: neither the log measure nor any of the three arrival bins is statistically significant (each $p>0.1$). A recent job arrival thus improves speed but has no detectable effect on quality.

Turning to cherry-picking, columns~(5)--(6) provide support for \hypref{H1b}: the log specification in column~(5) shows that cherry-picking decreases as time since the last arrival goes up (log-odds $-0.255$, $p<0.01$). The odds of cherry-picking decrease by about 16\% for each doubling of the time since the last arrival ($2^{-0.255} = 0.84$). That is, workers pick easier jobs following a new job arrival. The categorical specification in column~(6) shows the same pattern: relative to the $>$60 second baseline, the odds of cherry-picking are elevated after a recent arrival and fall as time since the last arrival increases (log-odds $1.312$, $1.368$, and $0.794$ for the $\leq$10, 10--30, and 30--60 second bins, all $p<0.01$). The two most recent bins are statistically indistinguishable from each other ($\leq$10 vs.\ 10--30 seconds; Wald test, $p=0.62$), while cherry-picking in the 10--30 second bin is significantly higher than in the 30--60 second bin ($p<0.001$).

\begin{table}[tb!]
\renewcommand{\arraystretch}{1.1}
\caption{Effect of Workload Salience on Performance and Cherry-Picking}
\label{tab:reg:H1}
\footnotesize
\centering
\begin{tabular}{lC{1.6cm}C{1.6cm}C{1.6cm}C{1.6cm}C{1.6cm}C{1.6cm}}
\toprule
\multicolumn{1}{r}{\textbf{Dependent Variable:}} & \multicolumn{2}{c}{\emph{Job Completion Time (sec)}} & \multicolumn{2}{c}{\emph{Job Error (0/1)}} & \multicolumn{2}{c}{\emph{Cherry-Picked (0/1)}} \\
\cmidrule(lr){2-3} \cmidrule(lr){4-5} \cmidrule(lr){6-7}
\multicolumn{1}{r}{\textbf{Sample:}} & \multicolumn{2}{c}{All rounds} & \multicolumn{2}{c}{Error-tolerant rounds} & \multicolumn{2}{c}{All rounds} \\
\cmidrule(lr){2-3} \cmidrule(lr){4-5} \cmidrule(lr){6-7}
& (1) & (2) & (3) & (4) & (5) & (6) \\ \midrule
\\[-1.5ex]
\emph{Log(Time Since Last Job Arrival + 1)} & 1.370*** &  & 0.036 &  & $-$0.255*** &  \\
 & (0.248) &  & (0.057) &  & (0.041) &  \\
\addlinespace
\emph{Last Job Arrival: $\leq$10 sec} &  & $-$4.714*** &  & $-$0.045 &  & 1.312*** \\
 &  & (0.881) &  & (0.222) &  & (0.217) \\
\addlinespace
\emph{Last Job Arrival: 10--30 sec} &  & $-$4.072*** &  & $-$0.066 &  & 1.368*** \\
 &  & (1.179) &  & (0.322) &  & (0.231) \\
\addlinespace
\emph{Last Job Arrival: 30--60 sec} &  & $-$1.643 &  & 0.254 &  & 0.794*** \\
 &  & (1.677) &  & (0.220) &  & (0.252) \\
\midrule
Estimator & OLS & OLS & Logit & Logit & Logit & Logit \\
Round FE & Yes & Yes & Yes & Yes & Yes & Yes \\
Job Type Controls & Yes & Yes & Yes & Yes & -- & -- \\
Queue Length Controls & Yes & Yes & Yes & Yes & Yes & Yes \\
Baseline Ability Controls & Yes & Yes & Yes & Yes & Yes & Yes \\
Observations & 4,200 & 4,200 & 1,396 & 1,396 & 3,379 & 3,379 \\
Subjects & 100 & 100 & 100 & 100 & 100 & 100 \\
Adj.\ $R^{2}$ & 0.156 & 0.155 & -- & -- & -- & -- \\
\bottomrule
\end{tabular}
\vspace{0.3em}

\begin{minipage}{0.97\textwidth}
\setlength{\baselineskip}{0.1\baselineskip}
\scriptsize \textit{Notes:} Cols.~(1)--(2) report OLS coefficients. Cols.~(3)--(6) report log-odds coefficients from logit regressions. Baseline ability controls are completion time and redo count in Round 0. Cols.~(3)--(4) use error-tolerant rounds, Medium and Large jobs (recall that no mistakes are allowed for Small jobs). Cols.~(5)--(6) include all decisions made when a cherry-picking opportunity was available. Standard errors clustered at the subject level in parentheses. The omitted job arrival category is $>60$ sec. Tests of directional hypotheses (\hypref{H1a} and \hypref{H1b}) use one-sided $p$-values; remaining comparisons use two-sided $p$-values. *** $p<0.01$, ** $p<0.05$, * $p<0.1$.
\end{minipage}
\end{table}

We next test \hypref{H2}, i.e., whether cherry-picking affects individual performance. Table~\ref{tab:reg:H2} regresses completion time (cols.~1--3, zero-tolerance rounds) and the job-error indicator (cols.~4--6, error-tolerant rounds restricted to Medium and Large jobs) on a \emph{Cherry-Picked} indicator that equals one when the worker selects a job easier than the one at the front of the queue and zero otherwise. All columns are restricted to selections made when a cherry-picking opportunity was available. Within each three-column block, column~(1)/(4) includes only round and job-type fixed effects; column~(2)/(5) adds baseline-ability controls; column~(3)/(6) replaces those with subject fixed effects.\footnote{A more granular analysis in which we decompose deviations from FIFO by type (picked easier, picked same difficulty level, or picked more difficult) is reported in Table~\ref{tab:reg:H2:direction} in \S\ref{ec:robustness}. Furthermore, Table~\ref{tab:reg:H2:sku} reruns the specification in Table~\ref{tab:reg:H2}  with controls for the individual items in each job. Both robustness checks support the conclusions drawn from Table~\ref{tab:reg:H2}.}

For completion time (cols.~1--3), cherry-picking is not associated with faster or slower job completion: the coefficient is small and insignificant in all three specifications. For errors (cols.~4--6), we find a robust positive association: cherry-picked jobs exhibit more errors (log-odds $0.61$ in both cols.~(4) and~(5), both significant at $p<0.05$), corresponding to an average marginal effect of $6.5$ percentage points.\footnote{Errors are almost always wrong substitutions: $84.5\%$ are incorrect items, $12.7\%$ are missing items, and $2.8\%$ are wrong quantities.} However, in the fixed-effects logit (column~6), which focuses on the effects of cherry-picking within-worker, the coefficient drops to $0.17$ and is no longer significant.\footnote{The fixed-effects logit in column~(6) is identified only from the workers whose error outcome varies across jobs (i.e., people who err at least once; here 54 of 100); subjects who never err drop out \citep[a general feature of nonlinear fixed-effects models, see][]{chamberlain1980}. The sample is further restricted to selections from mixed-difficulty queues, i.e., scenarios where a cherry-picking opportunity is available. For robustness, we replicate the analysis using a linear probability model with subject fixed effects, which retains all 100 workers. This analysis yields similar results (see Table~\ref{tab:reg:lpm} in \S\ref{ec:robustness}).} In other words, the positive relationship between cherry-picking and error rates is driven by \emph{which} workers cherry-pick rather than by the act of cherry-picking itself. 

\begin{table}[bt!]
\renewcommand{\arraystretch}{1.1}
\caption{Cherry-Picking and Performance}
\label{tab:reg:H2}
\footnotesize
\centering
\begin{tabular}{lC{1.5cm}C{1.5cm}C{1.5cm}C{1.5cm}C{1.5cm}C{1.5cm}}
\toprule
\multicolumn{1}{r}{\textbf{Dependent Variable:}} & \multicolumn{3}{c}{\emph{Job Completion Time (sec)}} & \multicolumn{3}{c}{\emph{Job Error (0/1)}} \\
\cmidrule(lr){2-4} \cmidrule(lr){5-7}
\multicolumn{1}{r}{\textbf{Sample:}} & \multicolumn{3}{c}{Zero-tolerance rounds} & \multicolumn{3}{c}{Error-tolerant rounds} \\
\cmidrule(lr){2-4} \cmidrule(lr){5-7}
& (1) & (2) & (3) & (4) & (5) & (6) \\
\midrule
\emph{Cherry-Picked} & 0.123 & 0.276 & $-$0.658 & 0.611** & 0.614** & 0.171 \\
 & (1.547) & (1.480) & (2.794) & (0.283) & (0.290) & (0.581) \\
\midrule
Estimator & OLS & OLS & OLS & Logit & Logit & FE Logit \\
Round FE & Yes & Yes & Yes & Yes & Yes & Yes \\
Subject FE & No & No & Yes & No & No & Yes \\
Baseline Ability Controls & No & Yes & No & No & Yes & No \\
Job Type Controls & Yes & Yes & Yes & Yes & Yes & Yes \\
Observations & 1,682 & 1,682 & 1,682 & 1,156 & 1,156 & 609 \\
Subjects & 100 & 100 & 100 & 100 & 100 & 54 \\
Adj.\ $R^{2}$ & 0.099 & 0.108 & 0.124 & -- & -- & -- \\
\bottomrule
\end{tabular}
\vspace{0.3em}

\begin{minipage}{0.88\textwidth}
\setlength{\baselineskip}{0.1\baselineskip}
\scriptsize \textit{Notes:} \emph{Cherry-Picked} equals one when the selected job is easier than the job at the front of the queue (an easy-first selection) and zero otherwise. OLS coefficients in cols.~(1)--(3); logit log-odds coefficients in cols.~(4)--(6), with col.~(6) a fixed-effects logit. Standard errors clustered at the subject level in parentheses. All columns restricted to selections made when a cherry-picking opportunity was available. Cols.~(1)--(3) use zero-tolerance rounds and all difficulties; cols.~(4)--(6) use error-tolerant rounds restricted to Medium and Large jobs (Small jobs have a zero error rate by design). One-sided $p$-values are reported for the pre-registered directional prediction (\hypref{H2}); the remaining comparisons use two-sided $p$-values. *** $p<0.01$, ** $p<0.05$, * $p<0.1$.
\end{minipage}
\end{table}

\begin{hyp}{Result 1}
Higher workload salience increases cherry-picking and improved speed (adjusted for job complexity). Workers complete jobs faster and are more likely to cherry-pick following a recent arrival, consistent with \hypref{H1a} and \hypref{H1b}. Error rates are not affected by workload.
\end{hyp}

\begin{hyp}{Result 2}
Cherry-picking is positively associated with errors, but the association is driven by selection rather than causation. Thus, \hypref{H2a} is supported while \hypref{H2b} is not. 
\end{hyp}

\subsubsection{Additional Analysis}\label{sec:exp1a:addl}
To better understand cherry-picking behavior and its implications we perform three additional tests. Overall, these tests suggest that high cherry-picking rates are a marker of careless execution rather than a result of a deliberate strategic trade-off or a marker of underlying ability. 

\begin{itemize}
\item[(i)] We have so far only examined completion times in the zero-tolerance round where errors are not permitted. However, we can also estimate the effects of cherry-picking on completion time in the \textit{error-tolerant }rounds (see Table~\ref{tab:reg:H2:time} for regression results). We find that cherry-picking is not associated with time savings during those rounds. This suggests that the increase in error rates is not a result of a strategic trade-off wherein error-prone workers finish sooner by cutting corners.  
\item[(ii)] Relatedly, cherry-picking rates are correlated with the workers' Round~0 (baseline) redo count ($\rho=0.27$, $p=0.007$; Spearman) but are uncorrelated with their Round~0 completion speed ($\rho=0.09$, $p=0.35$). That is, the workers who cherry-pick \textit{approach the task differently} than those who do not. In particular, cherry-pickers are just as fast as the rest, but make more attempts at each job suggesting that they are less careful, but not necessarily less capable. In Experiment 2 we will further characterize different worker types and will identify personality traits that predict cherry-picking behavior.  
\item[(iii)] The error association is also separate from the workload mechanism documented under \hypref{H1}: adding the time since the last job arrival to the regressions leaves the cherry-picking coefficient unchanged; as before, job arrival timing affects completion times but not error rates (see Table~\ref{tab:reg:H2:arrival} in \S\ref{ec:robustness}). Thus, cherry-picking and workload salience are separate channels: workload salience drives speed without affecting quality (\hypref{Result 1}), while cherry-picking marks worker types prone to errors (\hypref{Result 2}). In Experiment 1b we will further examine whether the selection mechanism persists when job sequencing decisions are fully separated from potential workload salience effects. 
\end{itemize}

\subsection{Experiment 1b: Pre-committed Sequencing}\label{sec:exp1b}
Experiment~1a shows that workers vary in how often they cherry-pick and that cherry-picking is associated with making more errors but does not causally lead to poor performance. One interpretation of this result is that cherry-picking is a stable preference for completing easy work first and that workers who are generally less careful have that preference. An alternative interpretation is that less careful workers resort to cherry-picking under pressure, i.e., that the tendency to cherry-pick is a more situational behavior rather than an innate preference. To test this, as well as to examine broader questions related to selection autonomy, we introduce a treatment in which workers are asked to choose a sequencing rule (FIFO or EF) at the start of each round and are then locked into their choice for the remainder of that round. %

\subsubsection{Experiment Design}

Experiment~1b was conducted on Prolific using the same recruitment criteria as Experiment~1a. Please see \S\ref{ec:prereg} for pre-registration details. The task, job types, arrival schedule, round structure, and within-subject manipulations (error tolerance, arrival rate) were identical to Experiment~1a. The only difference was that, at the start of each main round (prior to seeing any jobs), workers were asked to choose between two sequencing rules: FIFO or EF; the choice screen is reproduced in Figure~\ref{fig:precommit_choice} in \S\ref{ec:instructions}. Once they made their selection, the chosen rule was enforced for the remainder of the round; workers could not deviate from their sequence within the round. We will refer to this condition as \textit{Pre-commit}.

\subsubsection{Hypotheses}
As noted earlier, Experiment~1b has two pre-registered goals. First, by having workers commit to a sequencing rule before each round, it elicits their deliberate preference between FIFO and EF, rather than inferring it from job selections made on the fly. We have no strong prior on this preference: both FIFO (the natural default) and EF (the natural cherry-picking rule) have intuitive appeal, and Experiment~1a reveals substantial heterogeneity across workers in terms of their preferred sequences. Second, by locking the chosen rule for the round, Experiment~1b tests whether real-time flexibility (the ability to reorder jobs mid-round) itself affects performance. Since ordering does not causally affect performance (\hypref{Result 2}), locking workers into a single rule should not matter, and performance under \textit{Pre-commit} should match performance under \textit{Endog}. We therefore do not expect either treatment to be superior.

\begin{hyp}{H3 (Preference)}
There is no discernible preference between EF and FIFO.
\end{hyp}

\begin{hyp}{H4 (Ordering and Autonomy)}
Pre-commitment to a job sequence does not affect average performance relative to fully discretionary sequencing. 
\end{hyp}

\smallskip In addition to these two pre-registered hypotheses, the pre-commitment design gives us an opportunity to re-examine the selection mechanism driving \hypref{Result 2}. In particular, we have seen that cherry-picking in Experiment~1a is at least in part situational: it occurs more frequently right after a new job arrives (\hypref{Result 1}). This type of load-adaptive behavior cannot occur in Experiment~1b where easy-first sequencing reflects a more deliberate preference (as opposed to a more impulsive response to load or stress). If the selection mechanism identified in Experiment 1a reflects a stable preference, then we should also observe it in Experiment 1b. We therefore predict the following:

\begin{hyp}{Post-hoc Prediction (Based on Result 2)}
(i) There are more errors in rounds where a worker pre-commits to EF than where a worker pre-commits to FIFO; (ii) this effect is not causal and disappears once subject fixed effects are included.
\end{hyp}

\subsubsection{Results: Summary Statistics}
\begin{figure}[t!]
\centering
\caption{Performance Comparisons: \textit{Endog} vs.\ \textit{Pre-commit}}
\includegraphics[width=0.62\textwidth]{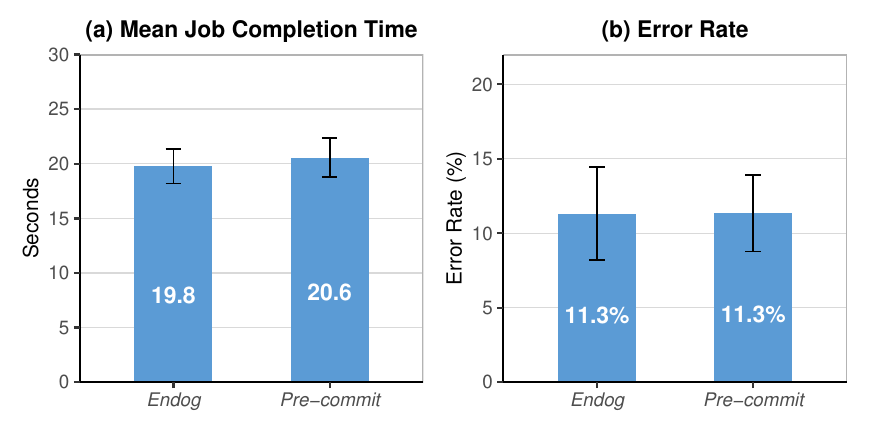}
\begin{minipage}{0.39\textwidth}\scriptsize
\textit{Notes:} Error bars are 95\% CIs, clustered at the subject level.
\end{minipage}
\label{fig:precommit}
\end{figure}

\begin{figure}[b!]
\centering
\caption{Job Completion Time and Error Rate by Chosen Rule (\textit{Pre-commit})}
\includegraphics[width=0.65\textwidth]{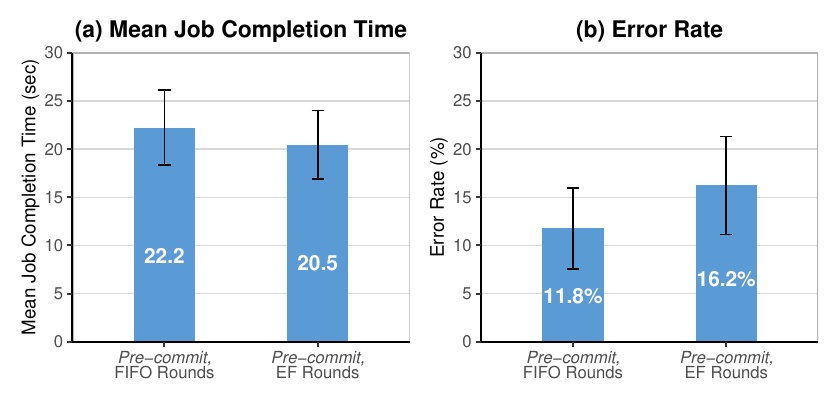}
\begin{minipage}{0.83\textwidth}\scriptsize
\textit{Notes:} \textit{Pre-commit} condition, Medium jobs only. Error bars are 95\% CIs, clustered at the subject level.
\end{minipage}
\label{fig:precommit_rule}
\end{figure}

Workers chose EF in $44.4\%$ of rounds and FIFO in the remaining $55.6\%$. This split is stable over time: the EF share is $44.1\%$, $44.1\%$, $45.9\%$, and $43.2\%$ in Rounds~1 through~4, with no discernible trend (Cuzick's trend test, $p=0.97$). Choices thus do not drift toward either rule as workers gain experience with the task. Figure~\ref{fig:precommit} compares aggregate performance between \textit{Endog} and \textit{Pre-commit}. The two conditions are nearly indistinguishable: completion times average $19.8$ vs.\ $20.6$ seconds (rank sum test, $p=0.98$) and the error rate is $11.3\%$ in both conditions (rank sum test, $p=0.49$), with widely overlapping confidence intervals.  Figure~\ref{fig:precommit_rule} is the analogue of Figure~\ref{fig:cp_task} for the \textit{Pre-commit} condition, again restricting to Medium jobs and splitting them by the rule the worker chose for the round (FIFO or EF).  Jobs completed under the EF rule are completed somewhat faster (20.5 vs.\ 22.2 seconds; rank sum test, $p=0.12$) and show a higher error rate than jobs completed under a chosen FIFO rule (16.2\% vs.\ 11.8\%; rank sum test, $p=0.07$). This is similar to the cherry-picking results observed in \textit{Endog}; however, the difference in error rates between EF and FIFO is about $46\%$ smaller than the corresponding difference in \textit{Endog} ($4.5$ vs.\ $8.3$ percentage points). As before, these pooled comparisons do not account for the repeated nature of the observations, which we address in the regression analysis below. %

\subsubsection{Hypothesis Tests}\label{sec:exp1b:tests}

We first test \hypref{H3} at the worker level. The mean subject-level EF share is $44.4\%$ and does not differ significantly from an even split (one-sample $t$-test against $50\%$, $p=0.14$). This suggests that there is no strong preference for either the EF or the FIFO rule in the population. Indeed, the distribution of choices is roughly bimodal: a majority of workers ($56.8\%$) choose the same rule in all four rounds ($34.2\%$ always FIFO, $22.5\%$ always EF), while the remaining $43.2\%$ mix the two rules.\footnote{We also check whether workers who consistently chose EF vs.\ those who consistently chose FIFO vs.\ those who mixed the two rules across rounds performed differently. We find that, of the three groups, workers who mixed had the highest error rates, though the differences are only marginally significant; \S\ref{ec:mixers} reports the analysis.}

\begin{hyp}{Result 3}
Workers show no strong aggregate preference between EF and FIFO, supporting \hypref{H3}.
\end{hyp}

We next test \hypref{H4} as well as the prediction that workers who select EF make more errors. Table~\ref{tab:reg:cprate_endog_precommit} pools observations from \textit{Endog} and \textit{Pre-commit}. In this table, the \emph{Easy-First} indicator equals one for cherry-picked \textit{Endog} jobs and for \textit{Pre-commit} jobs completed under a chosen EF rule. We include the \emph{Pre-commit} dummy to examine the treatment effect, as well as the interaction between  \emph{Easy-First} and \emph{Pre-commit} to examine whether the effect of job ordering differs between the two regimes. 

The results are as follows. First, the \textit{Pre-commit} dummy is small and not statistically significant in all specifications. Furthermore, the \emph{Easy-First} main effect continues to hold: cherry-picked jobs show higher error odds (log-odds $0.59$ and $0.55$, cols.~4--5, $p=0.025$ and $p=0.039$), corresponding to an average marginal effect of $5.9$ percentage points. However, if we evaluate the effect of easy-first ordering on errors within the \textit{Pre-commit} regime, it is positive but only marginally significant (col.~4: average marginal effect $+3.8$ pp; Wald test on the log-odds, $p=0.068$). This suggests that people who pre-commit to EF are somewhat less prone to errors than people who choose EF while performing the jobs. Indeed, in cols.~4--6, approximately 45\% of jobs are performed under EF in \textit{Pre-commit}, compared with 27\% in \textit{Endog}. The sharper sorting of error-prone workers into types under \textit{Endog} thus produces a sharper selection effect. As before, these selection effects disappear if we include subject fixed effects (col.~6).

\begin{table}[bt!]
\renewcommand{\arraystretch}{1.1}
\caption{Easy-First Behavior and Performance: \textit{Endog} vs.\ \textit{Pre-commit}}
\label{tab:reg:cprate_endog_precommit}
\footnotesize
\centering
\sbox{\tabnotebox}{\begin{tabular}{lC{1.5cm}C{1.5cm}C{1.5cm}C{1.5cm}C{1.5cm}C{1.5cm}}
\toprule
\multicolumn{1}{r}{\textbf{Dependent Variable:}} & \multicolumn{3}{c}{\emph{Job Completion Time (sec)}} & \multicolumn{3}{c}{\emph{Job Error (0/1)}} \\
\cmidrule(lr){2-4} \cmidrule(lr){5-7}
\multicolumn{1}{r}{\textbf{Sample:}} & \multicolumn{3}{c}{Zero-tolerance rounds} & \multicolumn{3}{c}{Error-tolerant rounds} \\
\cmidrule(lr){2-4} \cmidrule(lr){5-7}
& (1) & (2) & (3) & (4) & (5) & (6) \\
\midrule
\quad \emph{Endog} & [omitted] & [omitted] & [omitted] & [omitted] & [omitted] & [omitted] \\
\addlinespace
\quad \emph{Pre-commit} & $-$0.278 & $-$1.208 & [omitted] & 0.254 & 0.251 & [omitted] \\
 & (2.599) & (2.442) &  & (0.286) & (0.274) &  \\
\addlinespace
\quad \emph{Easy-First} & $-$0.069 & 0.145 & 0.944 & 0.589** & 0.553** & $-$0.463 \\
 & (2.780) & (2.787) & (5.293) & (0.300) & (0.314) & (0.422) \\
\addlinespace
\quad \emph{Pre-commit $\times$ Easy-First} & $-$1.296 & $-$0.478 & $-$1.826 & $-$0.241 & $-$0.206 & 0.713 \\
 & (3.653) & (3.522) & (6.174) & (0.382) & (0.391) & (0.630) \\
\midrule
Estimator & OLS & OLS & OLS & Logit & Logit & FE Logit \\
Round FE & Yes & Yes & Yes & Yes & Yes & Yes \\
Subject FE & No & No & Yes & No & No & Yes \\
Baseline Ability Controls & No & Yes & No & No & Yes & No \\
Observations & 1,437 & 1,437 & 1,437 & 1,448 & 1,448 & 637 \\
Subjects & 211 & 211 & 211 & 211 & 211 & 92 \\
Adj.\ $R^{2}$ & 0.048 & 0.058 & 0.098 & -- & -- & -- \\
\bottomrule
\end{tabular}}
\usebox{\tabnotebox}
\vspace{0.3em}
\par
\begin{minipage}{\wd\tabnotebox}
\setlength{\baselineskip}{0.1\baselineskip}
\scriptsize \textit{Notes:} Regressions of completion time (cols.~1--3, OLS) and a job error indicator (cols.~4--6, logit log-odds; col.~(6): fixed-effects logit), pooling \textit{Endog} and \textit{Pre-commit} jobs. All columns: Medium jobs only (No errors are allowed for Small jobs. Including Large jobs would change the meaning of the treatment indicator, since Large jobs cannot be cherry-picked in \textit{Endog} but can be present in a chosen-EF round in \textit{Pre-commit}.) Cols.~(1)--(3) use zero-tolerance rounds; cols.~(4)--(6) use error-tolerant rounds. \emph{Easy-First} = 1 if the job was cherry-picked (\textit{Endog}) or completed in a chosen-EF round (\textit{Pre-commit}). Standard errors clustered at the subject level in parentheses. The \emph{Pre-commit} dummy is constant within each worker and is absorbed by the subject fixed effects in cols.~(3) and~(6); it is therefore omitted there. One-sided $p$-values are reported for the pre-registered directional prediction (\hypref{H2}); the remaining comparisons use two-sided $p$-values. *** $p<0.01$, ** $p<0.05$, * $p<0.1$.
\end{minipage}
\end{table}

\begin{hyp}{Result 4}
Pre-commitment to a rule leaves performance unchanged relative to fully discretionary ordering, supporting \hypref{H4}. Consistent with the selection mechanism, easy-first ordering is associated with more errors, but we find no evidence that the effect is causal. Furthermore, selection effects are weaker than in Experiment~1a.
\end{hyp}

\subsection{Discussion}
The results of Experiment 1 replicate and unpack several prior findings in the literature, as well as identify a new mechanism that explains the relationship between job ordering and performance. First, the result that workers respond to perceived load by completing jobs faster and cherry-picking more frequently (\hypref{Result 1}) is consistent with the state-dependent service-rate literature \citep{edie1954,schultz1998,schultz1999,kc2009,delasay2019,kc2020}. Thus, our experimental task appears to reproduce key behaviors observed in field settings. Furthermore, we find that workload salience does not significantly affect quality, which is a more novel result. Second, consistent with the field evidence on discretionary task ordering \citep{ibanez2018,kc2020}, easy-first ordering is associated with worse performance (\hypref{Result 2}). All else equal, the chance of an error for a job that is completed out of FIFO sequence (ahead of more difficult jobs) is 66\% higher than when a job is completed in the order of arrival, while completion times are essentially unchanged.

What do these results add to the findings from the field? In field settings, a worker's task order, workload, and job mix are difficult to separate. We hold the set of jobs and their arrival sequence fixed: each worker faces the same jobs arriving on the same schedule. This allows us to compare how the same job is performed under different sequences and how the same worker performs when they cherry-pick vs.\ when they complete jobs in the order of arrival. We show that once worker fixed effects are included, the relationship between job ordering and performance essentially disappears. Thus, cherry-picking does not cause errors within a worker but is a marker of the type of worker that makes many errors. Indeed, additional analysis shows that the tendency to cherry-pick is visible before any queueing begins: cherry-picking correlates with redo counts in Round 0 (where the order was fixed), but is unrelated to Round 0 speed. This points to careless execution as a key driver of the errors, rather than to an underlying difference in abilities.\footnote{A further channel through which cherry-picking could hurt performance is browsing time: workers who deviate from FIFO may spend extra time scanning the job list \citep[a mechanism shown by][]{ibanez2018}. There is no evidence for this in our setting. High- and low-cherry-picking workers spend nearly identical time on the job list (median split, 23.7 vs.\ 22.4 seconds per round, $p=0.70$), and within worker, job-list time per job does not increase in rounds with more cherry-picking ($p=0.30$). A likely reason is that job difficulty is unambiguous in our setting: job types are clearly labeled by size, so workers can spot the easiest available job at a glance. In contrast, in the radiology setting of \citet{ibanez2018}, assessing case difficulty requires opening and reviewing images, which takes time and makes deviating from FIFO costly.}

Experiment~1b helps sharpen the selection mechanism. Preferences over job order appear to be stable worker traits. Individual rule choices are bimodal and persistent (\hypref{Result 3}), and the share of workers who always chose FIFO ($34\%$) is comparable to the share of Experiment~1a workers who never cherry-picked ($24\%$). Moreover, locking workers into their chosen rule does not affect performance (\hypref{Result 4}), so real-time flexibility itself neither helps nor hurts in our setting. Finally, the error gap between workers who chose EF and workers who chose FIFO in Experiment 1b is about half of the corresponding gap in Experiment~1a. The likely reason is that performing jobs in EF sequence is far more common in Experiment 1b than in Experiment~1a. Nearly half of all jobs are performed under a pre-committed EF rule in Experiment 1b, compared with about a quarter completed out of FIFO order in Experiment 1a. This suggests that some of the more careful workers also choose EF in Experiment 1b. In other words, the more deliberate rule choice in Experiment 1b sorts workers into types less sharply than cherry-picking in Experiment 1a, and weaker sorting produces a weaker selection effect.

Experiment~1 leads to three follow-up questions of managerial interest. First, what are the personality markers of the workers who cherry-pick and make errors? Second, is the discretion to choose one's job order helpful or harmful, i.e., would performance change if the firm imposed the sequence rather than the worker choosing it? Third, if workers speed up after seeing a job arrive, should arrivals be made more or less visible/salient? We next examine these questions in follow-up experiments.

\section{Experiment 2: Exogenous Job Sequencing}\label{sec:exp2}

To better understand how queue design affects behavior, we conduct a second experiment in which we impose a fixed job sequence on workers and examine its effects on performance, as well as conduct several personality tests to better characterize the selection mechanism identified in Experiment 1.

\subsection{Experiment 2a}\label{sec:exp2a}

In Experiment~2a we administer three between-subject treatments: a replication of the discretionary condition of Experiment~1a and two conditions in which the job sequence is imposed on the worker. 

\subsubsection{Experiment Design}

Experiment~2a was conducted on Prolific using the same recruitment criteria as Experiments~1a and~1b; workers who had participated in any earlier wave were not eligible. The task, job types, arrival schedules, round structure, compensation, and within-subject manipulations (error tolerance, arrival rate) were identical to Experiment~1a, with two changes.

First, before starting the task, participants completed a short battery of personality measures: the ten-item Big Five Inventory \citep[BFI-10;][]{rammstedt2007}, the fifteen-item Need for Closure scale \citep[NFCS-15;][]{roets2011}, and an eleven-point risk-appetite question \citep{dohmen2011};\footnote{To save time, we elicit risk appetite with the single general risk question rather than an incentivized elicitation such as the multiple-price-list approach \citep{holt2002}. \citet{dohmen2011} validate the self-reported measure that we use against an incentivized lottery experiment with real stakes and show that it reliably predicts incentivized risk taking. In Experiment~1 we administered the same battery of personality tests at the end of the session (as opposed to Experiment 2 where it was done at the beginning of the session). Eliciting the measures before the task avoids two concerns with end-of-session responses: participants may be fatigued, and their answers may be affected by how they performed in the main task.} all items are reproduced verbatim in \S\ref{ec:instruments}. These measures allow us to characterize worker types and to link personality to behavior in the experiment, i.e., job-order preferences, speed, and the quality of submitted work.  Each instrument targets a mechanism suggested by Experiment~1. The BFI-10 covers the Big Five, including conscientiousness, the trait plausibly tied to the careless execution driving the error results of Experiment~1. The Need for Closure scale captures discomfort with ambiguity, with subscales for decisiveness and preference for structure; workers high on these traits may find a queue of pending jobs aversive and gravitate toward easy jobs, which provide quick completions.  Finally, risk appetite may speak to the quality of submissions in the error-tolerant rounds, where submitting a job without ensuring its complete correctness is a small risky gamble since a single error allows one to proceed to the next task, while multiple errors require a redo. Workers with a higher risk appetite may thus check less and accept a higher error rate.

Second, participants were randomly assigned to one of three conditions:
\begin{enumerate}[nosep]
    \item \textit{Endog}: replication of Experiment~1a;
    \item \textit{Exog-FIFO}: jobs had to be completed in arrival order;
    \item \textit{Exog-EF}: jobs had to be completed from easiest to hardest 
\end{enumerate}
The pre-registered target sample was 300 participants (100 per condition). A total of 322 subjects completed the experiment with a final sample of 309 participants after all exclusions (see \S\ref{ec:prereg} for pre-registration details, detailed sample calculations, demographics, exclusions etc.).

\subsubsection{Hypotheses}
Our previous results (\hypref{Result 2} and \hypref{Result 4}) suggest  that  job ordering does not causally affect performance, and therefore putting a worker into an exogenously imposed job sequence (FIFO or EF) should not matter either. In contrast, \citet{ibanez2018} find that radiologists who deviate from the default queue order, especially by completing easy cases first, work more slowly; they estimate that removing worker discretion would have increased annual profits by roughly 3\%. \citet{kc2020} find similar effects in a hospital setting. Both papers conclude that limiting worker autonomy over task ordering could improve performance. Whether restricting autonomy would improve performance is, however, a separate question from whether the sequence matters. This is because restricting autonomy changes \textit{who} sets the job order, and neither our Experiment 1 nor prior literature varies this. Nonetheless, several studies in non-service related settings \citep[][see \textsection 2.3 for details]{bray2016,kagan2018,rusou2020} have shown that workers often underperform when choosing their own workflow. We therefore hypothesize:

\begin{hyp}{H5 (Exogenous Ordering)}
Exogenously imposed job ordering improves performance relative to discretionary job ordering.
\end{hyp}

\smallskip In addition to \hypref{H5}, our pre-registration specifies three secondary analyses. First, we examine whether the effect of imposed job sequence is specific to one of the two regimes (\textit{Exog-FIFO} vs.\ \textit{Exog-EF}). In particular, it is possible that an EF sequence helps weaker workers build routine on simple jobs before facing more complex ones. Second, we examine whether the effect of imposed ordering on performance is moderated by ability. Third, we examine whether any of the personality measures correlate with performance and/or with sequencing behavior.

\subsubsection{Results: Summary Statistics}

\begin{figure}[b]
\centering
\caption{Experiment 2a: Summary Statistics by Condition}
\label{fig:exp2_summary}
\includegraphics[width=0.65\textwidth]{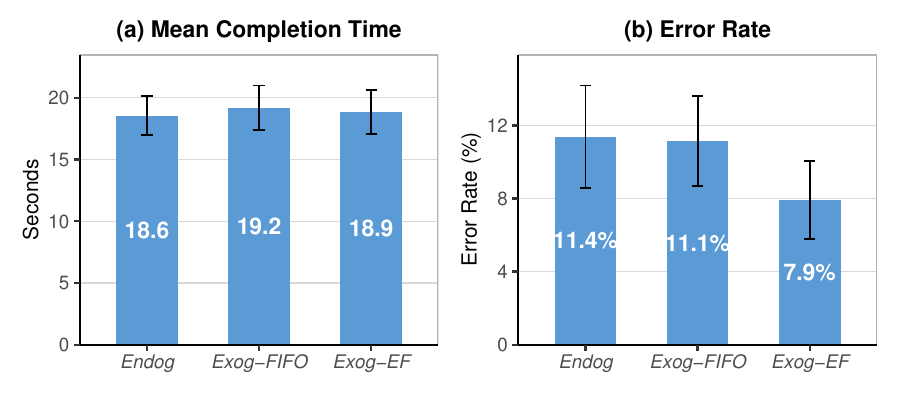}
\begin{minipage}{0.45\textwidth}\scriptsize
\textit{Notes:} Error bars are 95\% CIs, clustered at the subject level.
\end{minipage}
\end{figure}

Figure~\ref{fig:exp2_summary} presents mean completion times and error rates by condition. Completion times are nearly identical across the three arms: $18.6$, $19.2$, and $18.9$ seconds in \textit{Endog}, \textit{Exog-FIFO}, and \textit{Exog-EF}, with all pairwise rank sum tests insignificant ($p \geq 0.62$). However, there are some treatment differences in error rates. \textit{Endog} and \textit{Exog-FIFO} are statistically indistinguishable ($11.4\%$ vs.\ $11.1\%$; rank sum test, $p=0.88$), while \textit{Exog-EF} is roughly $30\%$ lower ($7.9\%$; $p=0.081$ vs.\ \textit{Endog} and $p=0.065$ vs.\ \textit{Exog-FIFO}). These raw comparisons provide some initial evidence that imposing a sequence does not slow workers down, and that the easiest-first sequence may improve quality.

\subsubsection{Hypothesis Tests}

Table~\ref{tab:reg:H5} reports regression analysis with completion time (cols.~1--2) and errors (cols.~3--4) as dependent variables on treatment dummies (\textit{Endog} is the omitted category). Columns~(2) and~(4) add the pre-registered interactions with baseline ability, which we discuss in the next subsection. Completion times do not differ across conditions: both treatment coefficients are small and insignificant (col.~1). For errors, imposing FIFO leaves the error rate unchanged (col.~3), while imposing EF reduces the odds of an error (log-odds $-0.426$, col.~3, $p=0.009$); the implied average marginal effect is $-3.3$ percentage points, corresponding to a reduction of roughly $29\%$ relative to the \textit{Endog} mean of $11.4\%$. The two imposed regimes also differ from each other: the EF--FIFO contrast is $-0.39$ in log-odds ($-3.1$ percentage points; Wald test, $p=0.021$). Thus the improvement predicted by \hypref{H5} is supported for the quality dimension and only for a specific regime: the EF sequence reduces error rates, while imposed FIFO leaves performance unchanged.\footnote{Given that errors drop sharply after the first round (see Figure~\ref{fig:summary}), we checked whether the reduction in error rates achieved by the \textit{Exog-EF} regime is only observed at the beginning of the experiment when workers are familiarizing themselves with the task. Contrary to that, we find that the error reduction is present in every round: relative to \textit{Endog}, \textit{Exog-EF} lowers the error rate by about $21\%$, $24\%$, $30\%$, and $51\%$ in Rounds~1 through~4, respectively. The differences in treatment effects across rounds are not statistically significant (treatment-by-round interactions are jointly insignificant, Wald test $p=0.51$, and a linear round trend is insignificant, $p=0.75$).}

\begin{hyp}{Result 5}
Imposing a job sequence does not affect completion times. The easiest-first sequence reduces error rates by about $30\%$ relative to discretionary ordering, supporting \hypref{H5} on the quality dimension; imposed FIFO leaves performance unchanged.
\end{hyp}

\begin{table}[bt!]
\renewcommand{\arraystretch}{1.1}
\caption{Treatment Effects on Task Performance}
\label{tab:reg:H5}
\footnotesize
\centering
\begin{tabular}{lC{1.9cm}C{1.9cm}C{1.9cm}C{1.9cm}}
\toprule
\multicolumn{1}{r}{\textbf{Dependent Variable:}} & \multicolumn{2}{c}{\emph{Job Completion Time (sec)}} & \multicolumn{2}{c}{\emph{Job Error (0/1)}} \\
\cmidrule(lr){2-3} \cmidrule(lr){4-5}
\multicolumn{1}{r}{\textbf{Sample:}} & \multicolumn{2}{c}{Zero-tolerance rounds} & \multicolumn{2}{c}{Error-tolerant rounds} \\
\cmidrule(lr){2-3} \cmidrule(lr){4-5}
& (1) & (2) & (3) & (4) \\
\midrule
\emph{Endog} & [omitted] & [omitted] & [omitted] & [omitted] \\
\addlinespace
\emph{Exog-FIFO} & 0.670 & 0.576 & $-$0.037 & $-$0.032 \\
 & (1.141) & (1.110) & (0.157) & (0.158) \\
\addlinespace
\emph{Exog-EF} & $-$0.005 & $-$0.021 & $-$0.426*** & $-$0.402** \\
 & (1.128) & (1.103) & (0.180) & (0.181) \\
\addlinespace
\emph{Baseline Ability (per SD)} & $-$1.829*** & $-$3.368*** & $-$0.092** & $-$0.132** \\
 & (0.662) & (0.918) & (0.046) & (0.059) \\
\addlinespace
\emph{Exog-FIFO $\times$ Baseline Ability} &  & 3.528*** &  & 0.009 \\
 &  & (1.295) &  & (0.115) \\
\addlinespace
\emph{Exog-EF $\times$ Baseline Ability} &  & 1.684 &  & 0.197* \\
 &  & (1.295) &  & (0.114) \\
\midrule
Estimator & OLS & OLS & Logit & Logit \\
Round FE & Yes & Yes & Yes & Yes \\
Job Type Controls & Yes & Yes & Yes & Yes \\
Observations & 6,489 & 6,489 & 4,326 & 4,326 \\
Subjects & 309 & 309 & 309 & 309 \\
Adj.\ $R^{2}$ & 0.120 & 0.122 & -- & -- \\
\bottomrule
\end{tabular}
\vspace{0.3em}

\begin{minipage}{0.79\textwidth}
\setlength{\baselineskip}{0.1\baselineskip}
\scriptsize \textit{Notes:} Cols.~(1)--(2): OLS coefficients reported, zero-tolerance rounds, all job types. Cols.~(3)--(4): Logit, log-odds reported, error-tolerant rounds restricted to Medium and Large jobs (Small jobs have a zero error rate by design). \emph{Baseline Ability}: reverse-coded composite of standardized Round~0 time per job and redo count (per SD; higher = faster and more accurate at baseline); treatment coefficients show effects at average baseline ability. Standard errors clustered at the subject level in parentheses. One-sided $p$-values are reported for directional hypotheses; the remaining comparisons use two-sided $p$-values. *** $p<0.01$, ** $p<0.05$, * $p<0.1$.
\end{minipage}
\end{table}

\begin{figure}[bt!]
\centering
\caption{Treatment Effects by Baseline Ability}
\includegraphics[width=0.85\textwidth]{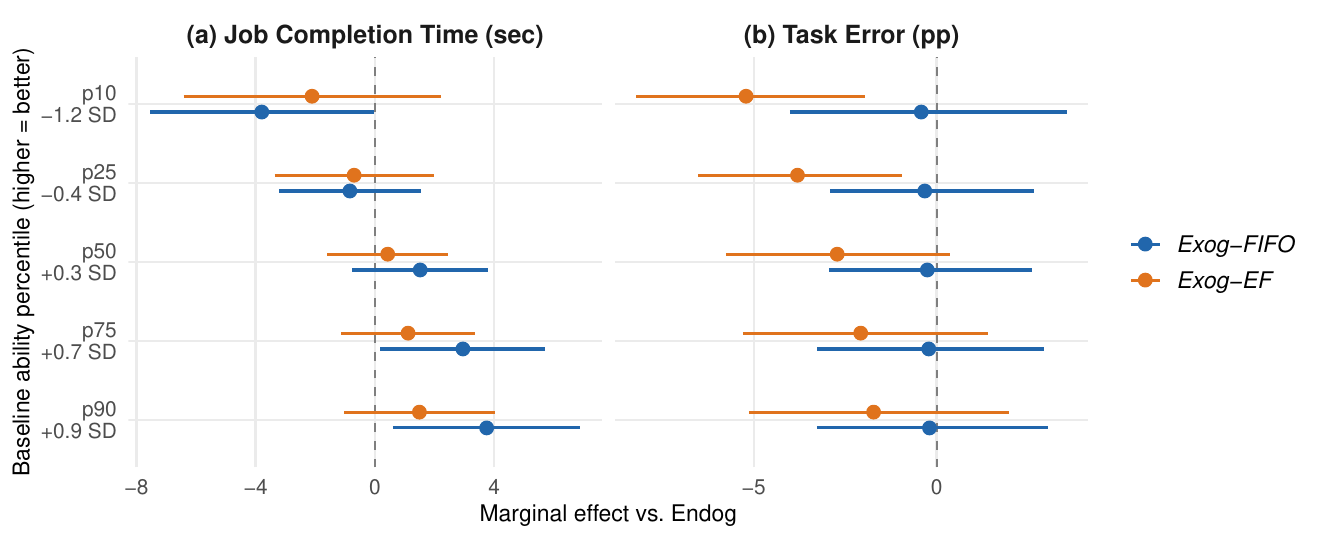}
\begin{minipage}{0.76\textwidth}\scriptsize
\textit{Notes:} Marginal effects based on cols.~(2) and~(4) of Table~\ref{tab:reg:H5}. Error bars are 95\% CIs.
\end{minipage}\label{fig:treatment_margins}
\end{figure}

In addition to average treatment effects, it is informative to examine heterogeneity across worker ability levels. Columns~(2) and~(4) of Table~\ref{tab:reg:H5} interact the treatment dummies with baseline ability. Figure~\ref{fig:treatment_margins} plots the marginal effects at different percentiles of ability with the lowest (highest) ability being the 10th (90th) percentile. Several observations are in order. First, imposing a sequence compresses the relationship between baseline ability and speed. In particular, column 2 shows that greater baseline ability is associated with better performance on the main task ($p<0.001$), but under \textit{Exog-FIFO} this effect disappears almost entirely (interaction $p=0.006$; effect of baseline ability not significant under \textit{Exog-FIFO}, $p=0.83$). That is, under \textit{Exog-FIFO}, more capable workers lose the speed advantage they would otherwise enjoy from being able to control the job sequence, while slow workers catch up. Indeed, panel (a) of Figure~\ref{fig:treatment_margins} shows that the weakest workers (10th percentile) are marginally faster under \textit{Exog-FIFO} than under \textit{Endog} ($p=0.058$), while the most capable workers (90th percentile) are significantly slower ($p=0.022$).  Second, the error reduction from \textit{Exog-EF} declines with ability, though only marginally ($p=0.083$). In particular, the treatment effect of \textit{Exog-EF} is $-2.7$ percentage points at the median of baseline ability and $-5.2$ percentage points at the 10th percentile. The gains from imposed EF are thus concentrated among weaker workers. In contrast, high-ability workers (75th percentile and above) make few errors regardless of the job sequence (panel (b) of Figure~\ref{fig:treatment_margins}).

\subsubsection{Personality Correlates}\label{sec:exp2a:pers}
Our final pre-registered analysis examines which personality measures predict behavior, in particular, the tendency to pick jobs out of order and the two performance components, speed and quality. Tables~\ref{tab:cormatrix} (full sample) and~\ref{tab:cormatrix:endog} (\textit{Endog} sample) in the e-companion report the full correlations among personality and demographic measures and performance outcomes. Most measures are weak predictors: across the Big Five, the need-for-closure scales, and the demographic variables, correlations with cherry-picking, speed, and errors are small and mostly not statistically significant. Three correlations, however, stand out. First, preference for structure is the only measure that predicts cherry-picking ($r=0.21$, $p=0.03$): workers who find an unstructured queue aversive gravitate toward quick, easy completions. Second, extraversion correlates with both errors ($r=0.30$) and completion times ($r=0.32$). Third, risk appetite correlates with errors ($r=0.28$). We will next examine how these traits map onto worker types defined jointly by job ordering behavior and performance. 

\begin{figure}[tb]
\centering
\caption{Trait Profiles of Worker Types (Experiment 2a, \textit{Endog} Treatment)}
\includegraphics[width=\textwidth]{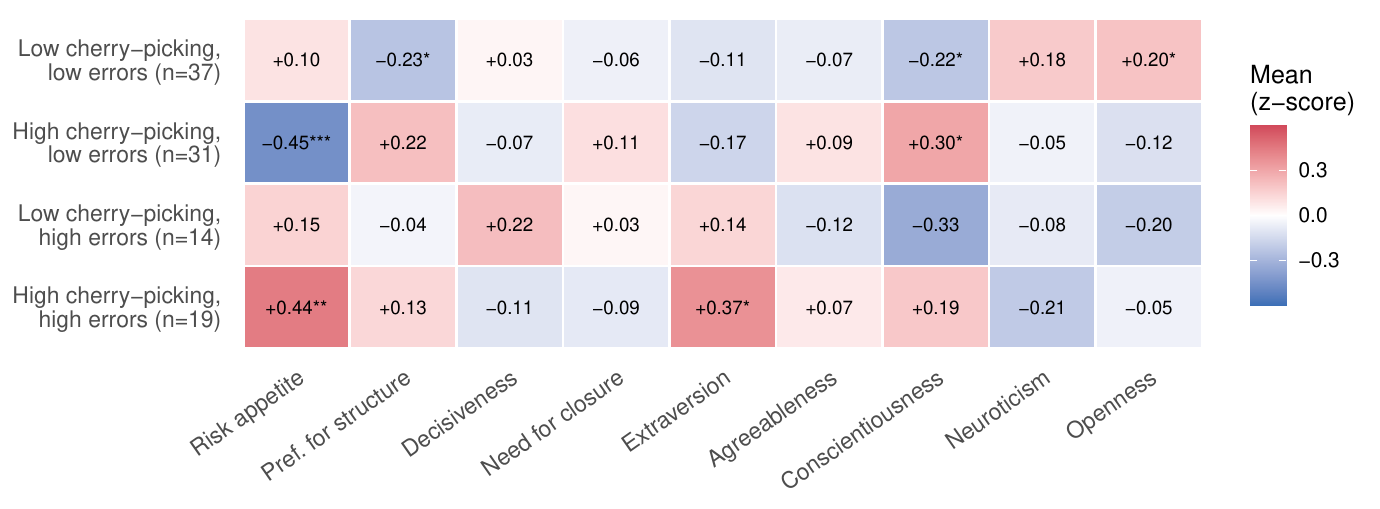}
\begin{minipage}{0.95\textwidth}\scriptsize
\textit{Notes:} Types are defined by whether a worker is above (``high'') or below (``low'') the median on the cherry-picking rate and on the error rate; the cherry-picking rate is the share of cherry-picking opportunities on which the worker took the easier job. Asterisks denote a difference from the rest of the sample in a rank sum test: *** $p<0.01$, ** $p<0.05$, * $p<0.1$.
\end{minipage}
\label{fig:personality}
\end{figure}

In Figure~\ref{fig:personality} we split workers at the median cherry-picking rate and the median error rate and report the average standardized trait scores of the four resulting types. Workers who combine frequent cherry-picking with frequent errors score higher than the rest of the sample on risk appetite ($+0.44$~SD, rank sum test $p=0.018$) and, more weakly, on extraversion ($+0.37$~SD, $p=0.082$). By contrast, cherry-pickers who make few mistakes are risk-averse ($-0.45$~SD, $p=0.003$). A natural interpretation of these effects is that in the error-tolerant rounds, submitting a job without double-checking all the items is risky (since making more than one error requires a redo), and workers with a higher tolerance for risk check less. Notably, this gamble does not help improve speed: risk appetite and error rate are each uncorrelated with completion time ($r=0.02$, $p=0.83$, and $r=0.04$, $p=0.72$), which is consistent with our interpretation that cherry-picking and high error rates are not caused by low ability, but rather by careless execution.

\begin{hyp}{Result 6}
Greater risk appetite is a marker of the worker type that makes more errors and cherry-picks easier jobs from the queue.  
\end{hyp}

\subsection{Experiment 2b: Invisible Queue}\label{sec:exp2b}

In all preceding experiments (1a, 1b, and 2a), workers see the full queue: they observe every job arrival, its difficulty, and its position in the queue. Experiment~2b removes this visibility while holding job sequences exactly the same as in the \textit{Exog-FIFO} treatment in Experiment 2a.

\subsubsection{Experiment Design}
Workers in Experiment~2b receive jobs in FIFO order but see only the current job; they do not observe the number of remaining jobs, their difficulty, or when new jobs arrive. We refer to this condition as \textit{Invisible-FIFO}. Job types, round structure, and within-subject manipulations (mistake tolerance, arrival rate) are the same as in Experiments~1a, 1b, and~2a. When the current job is completed, the next job in FIFO order is revealed. See \S\ref{ec:prereg} for pre-registration, sample size, demographics and exclusion details.

\subsubsection{Hypotheses}
Prior literature and our own results suggest that visibility may affect performance. In particular, \hypref{Result 1} suggests that workers work faster immediately after a new job arrives in the queue, which, by definition, requires that the worker is able to observe the queue and the arrivals. More broadly, the behavioral operations literature has documented that visible backlogs create time pressure that motivates faster work \citep{schultz1998, kc2009, delasay2019}. Furthermore, \citet{shunko2018} show that restricting queue-length visibility slows down servers, although they focus on a multi-server setting where social loafing plays a big role. If workers in our setting use queue information to calibrate effort or pace themselves, then removing visibility could reduce performance. 

\begin{hyp}{H6 (Visibility)}
Workers in the \textit{Invisible-FIFO} condition exhibit lower performance than workers in the \textit{Exog-FIFO} condition.
\end{hyp}

Arguments against queue visibility come from the psychology literature on worker motivation and cognitive load. Uncompleted jobs waiting ahead tend to stay cognitively active and can intrude on the work currently being performed \citep{zeigarnik1927, masicampo2011}, so a worker who sees difficult jobs waiting may worry about them while completing the current one. If the queue pulls attention away from the job at hand, removing queue information could improve performance.

\subsubsection{Results: Summary Statistics}

\begin{figure}[b]
\centering
\caption{Experiment 2b: Summary Statistics by Condition}
\label{fig:exp2b_summary}
\includegraphics[width=0.60\textwidth]{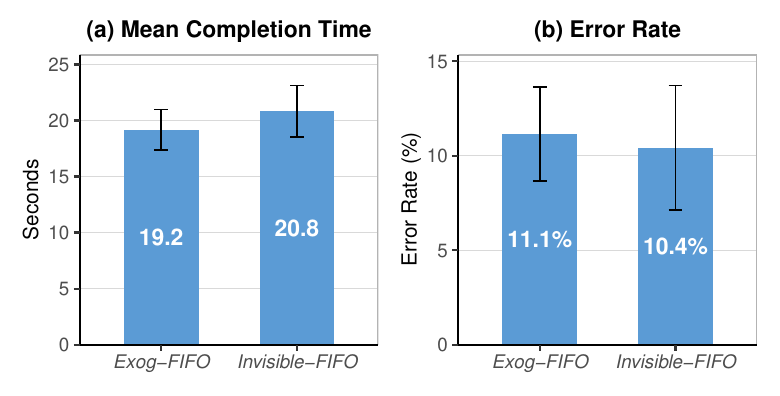}
\begin{minipage}{0.42\textwidth}\scriptsize
\textit{Notes:} Error bars are 95\% CIs, clustered at the subject level. 
\end{minipage}
\end{figure}

Figure~\ref{fig:exp2b_summary} presents mean completion times and error rates in the \textit{Invisible-FIFO} condition ($n=100$) alongside the \textit{Exog-FIFO} arm of Experiment~2a ($n=107$). The two conditions are statistically indistinguishable on both dimensions: completion times average $20.8$ vs.\ $19.2$ seconds (rank sum test, $p=0.25$) and error rates are $10.4\%$ vs.\ $11.1\%$ (rank sum test, $p=0.40$). These raw comparisons suggest that removing queue visibility neither slows workers down nor affects the quality of their work.

\subsubsection{Hypothesis Tests}

Table~\ref{tab:reg:H6} reports regression coefficients: completion time (col.~1) and the job-error indicator (col.~2) on an \textit{Invisible-FIFO} dummy, with \textit{Exog-FIFO} as the omitted category. Neither performance outcome is significantly affected by the treatment. The completion-time coefficient is small and not statistically significant ($1.261$ seconds, col.~1, $p=0.18$), and the error coefficient is essentially zero (log-odds $0.003$, col.~2, $p=0.49$). \hypref{H6} is therefore not supported: removing the queue display, and with it the urgency of seeing new arrivals that drive the load-adaptive behavior in \hypref{Result 1}, does not affect average performance in the long term.

\begin{hyp}{Result 7}
Removing visibility leaves completion times and error rates unchanged. \hypref{H6} is not supported.
\end{hyp}

\begin{table}[tb!]
\renewcommand{\arraystretch}{1.1}
\caption{Queue Visibility and Task Performance}
\label{tab:reg:H6}
\footnotesize
\centering
\begin{tabular}{lC{4cm}C{4cm}}
\toprule
\multicolumn{1}{r}{\textbf{Dependent Variable:}} & \emph{Job Completion Time (sec)} & \emph{Job Error (0/1)} \\
\cmidrule(lr){2-2} \cmidrule(lr){3-3}
\multicolumn{1}{r}{\textbf{Sample:}} & Zero-tolerance rounds & Error-tolerant rounds \\
\cmidrule(lr){2-2} \cmidrule(lr){3-3}
& (1) & (2) \\
\midrule
\emph{Exog-FIFO} & [omitted] & [omitted] \\
\addlinespace
\emph{Invisible-FIFO} & 1.261 & 0.003 \\
 & (1.400) & (0.219) \\
\midrule
Estimator & OLS & Logit \\
Round FE & Yes & Yes \\
Job Type Controls & Yes & Yes \\
Baseline Ability Controls & Yes & Yes \\
Observations & 4,362 & 2,888 \\
Subjects & 207 & 207 \\
Adj.\ $R^{2}$ & 0.071 & -- \\
\bottomrule
\end{tabular}
\vspace{0.3em}

\begin{minipage}{0.74\textwidth}
\setlength{\baselineskip}{0.1\baselineskip}
\scriptsize \textit{Notes:} Regressions pooling the \textit{Exog-FIFO} arm of Experiment~2a and the \textit{Invisible-FIFO} condition of Experiment~2b, omitted category is \textit{Exog-FIFO}. Col.~(1): completion time, zero-tolerance rounds, all job types (OLS). Col.~(2): job error indicator, error-tolerant rounds, Medium and Large jobs (logit, log-odds reported). Standard errors clustered at the subject level in parentheses. *** $p<0.01$, ** $p<0.05$, * $p<0.1$.
\end{minipage}
\end{table}

\subsection{Discussion}
A practical reading of the field evidence on discretionary task ordering is that managers should limit worker discretion or actively guide workers' task selection \citep{ibanez2018,kc2020}. Experiment~2a supports this recommendation. Indeed, our results suggest that not only \textit{whether} the firm removes discretion matters, but also \textit{what} sequence it imposes. Imposing FIFO, which is often the default rule in practice, does not affect performance. In contrast, imposing easy-first reduces errors by roughly $30\%$ (\hypref{Result 5}). Unlike Experiment~1, where easy-first ordering is a marker of careless workers (rather than a cause of their errors), imposing the easy-first sequence throughout the work period leads to a gradual increase in job complexity, which, as our data show, tends to be beneficial for work quality. 

These gains and costs are not evenly distributed across ability levels. The quality gains from an imposed easiest-first sequence affect low performers, while the reduction in speed due to an imposed sequence affects the strongest workers, who lose the advantage of managing their own queue. Imposed structure therefore helps the workers who need support and holds back the workers who do not. This result suggests that an effective work design should respond to worker seniority and performance: default structured sequences, with complexity increasing over the shift, for new or low-performing workers, and sequencing discretion as earned autonomy for top performers.

The personality trait analysis further characterizes worker types who are both error-prone and often choose easy-first sequences. In particular, the most robust marker of these workers is greater risk appetite, which correlates with both error rates and easy-first sequencing (\hypref{Result 6}), but not with speed. This is consistent with the interpretation that these workers are not trading accuracy for speed, nor are they generally less capable performers. Rather, these workers simply check their work less and rely more on chance.

Finally, the null result for queue visibility (\hypref{Result 7}) adds to the behavioral literature examining worker effort in multi-server settings. In these settings, queue visibility can make the workload salient and speed servers up \citep{shunko2018} and can determine whether they try to cooperate or free ride on each other's effort \citep{rosokha2024}. In our single-worker setting with a fixed sequence, the social loafing motive is absent, and hiding the queue leaves both speed and quality unchanged. The short-term productivity bursts thus do not translate into aggregate performance gains, suggesting that firms cannot rely on queue visibility as a lever to change individual worker performance.

\section{Concluding Remarks}\label{sec:conclusions}

Queue-based work is the norm in many service settings: warehouse workers fill orders from a list, radiologists pick cases, and back-office workers select the order in which they process claims and requests. To design such workflows well, firms need to understand how workers use the discretion to arrange the jobs and what happens when this discretion is taken away. In this paper, we studied queue design by developing an experimental approach in which every worker faces the same jobs arriving on the same schedule and in which speed and quality are measured separately at the item level, and by examining how job sequence, worker autonomy, and queue visibility affect performance.

\subsubsection*{Summary of Results}
Field studies of discretionary ordering show that when workers deviate from FIFO to browse and complete easy jobs first, throughput can decrease, in part because inspecting the queue to decide what to do next takes away from their time and attention \citep{ibanez2018,kc2020}. We add to this research by incorporating separate measures of quality and speed into performance measurement, and by identifying and characterizing a new mechanism based on worker self-selection. In particular, we show that workers who perform easy tasks first make more errors, but that within worker the effect disappears. Thus, easy-first job ordering can be a marker of less careful workers. Imposing FIFO does not affect average speed or quality, while imposing easy-first reduces errors by approximately $30\%$, with the gains concentrated among weaker performers. Furthermore, autonomy over job sequencing may help top performers improve speed. Finally, hiding the queue leaves average performance unchanged, even though recent arrivals produce short-term productivity bursts.

\subsubsection*{Practical Implications}
For managers of queue-based service work, our results have three practical takeaways. First, our result that cherry-picking (completing jobs in easy-first sequence) marks error-prone workers and that this worker type scores highly on risk-taking suggests worker screening as a potential remedy. Employers in service industries, such as call centers and back-office processing, already screen applicants with job tests that include personality questions, and these tests have been shown to improve the quality of hires \citep{autor2008, hoffman2018}. Our results suggest that risk preference assessment may be a useful addition to such tests.\footnote{Simple behavioral measures of risk-taking already exist \citep[e.g.,][]{lejuez2002}, and are used in some online hiring assessments, but mostly for white-collar job recruiting, especially in banking and finance. See, e.g., \url{https://www.efinancialcareers.com/news/2019/08/jpmorgan-pymetrics}. Our findings suggest these measures would be valuable for screening service workers as well.} Second, the specific sequence the firm uses matters for the quality of work. Changing the default from FIFO to easiest-first is essentially free, and in our data it helps reduce errors, particularly for weaker workers. Flexible, discretionary workflow should be treated as earned autonomy for top performers or the most experienced workers. Third, managers should not expect queue visibility to affect individual productivity: any immediate urgency added by a visible queue (or by new job arrival announcements) appears to soften in the long term. Managers should, however, keep in mind that this result need not carry over to team settings, where a shared queue can change effort through comparison with coworkers.

\subsubsection*{Limitations and Extensions}
To keep the experiments focused, we deliberately removed from the experimental setup several features of service work that are relevant in practice. Our sessions are short (30 minutes, on average), jobs are independent, difficulty (job size) is clearly labeled, and errors are objectively verifiable; workers face no customers, no interruptions, and no consequences beyond their pay. Settings in which assessing job difficulty is costly may restore the browsing-time mechanism proposed by \citet{ibanez2018}. Longer horizons may lead to more fatigue or learning dynamics. The personality results are exploratory and correlational; further work with finer measurements (e.g., the full Big Five inventory instead of our shortened version)  can help address this. Incentivized preference elicitations, such as the multiple-price-list task of \citet{holt2002} or the ordered-lottery task of \citet{eckel2008}, could help measure risk and other types of preferences (e.g., loss aversion or intertemporal choice) more granularly. Further extensions include team settings or endogenous arrival processes in which worker speed and quality affect workload.

\subsubsection*{Outlook}
Our experiments are an early attempt to begin understanding the effects of queue design on server behavior, and many questions remain open. First, because structure helps weaker workers but slows down the strongest, sequences that adapt to the individual worker could be a viable technological solution to worker heterogeneity. A natural form of such adaptation is algorithmic guidance: \citet{bastani2025} show that machine-generated tips can improve workers'  sequencing decisions while adjusting to the worker's strategies and capabilities. A similar adaptive approach can be used to manage workflow in service settings. Second, any sequencing policy is also a priority rule from the customers' perspective: an easiest-first queue that improves quality makes the most complex cases wait the longest. This is especially important given that people are sensitive not only to the overall duration of a wait but also to their position in line \citep{buell2021} and to the fairness of the service order \citep{larson1987, althenayyan2025}. Finally, autonomy is an important driver of job satisfaction \citep{hackman1976, deci2000}, so sequencing and information sharing policies may affect morale and turnover. As technology continues to make it increasingly possible for managers to control what workers see and in what order they perform tasks, studies of worker and customer responses to these technologies can help us understand best practices. 

\bibliographystyle{informs2014} %
\bibliography{literature} %

\newpage \clearpage \raggedbottom

\ECSwitch

\ECHead{Electronic Companion}
\small

\noindent This companion includes the following:
\smallskip

\begin{itemize}[leftmargin=1.5cm]
    \item Pre-registration summary (\S\ref{ec:prereg})
    \item Pilot results (\S\ref{ec:pilot})
    \item Experimental instructions (\S\ref{ec:instructions})
    \item Additional analyses and robustness checks (\S\ref{ec:robustness})
\end{itemize}
\bigskip

\section{Pre-registrations}
\label{ec:prereg}
\FloatBarrier

All experiments in the main text were pre-registered on AsPredicted prior to data collection. All experiments recruited US-based Prolific workers with a 99\%+ approval rate and ages 18--40, and workers who had taken part in an earlier wave were not eligible for later ones. The age restriction was introduced after the pilot, where a small number of substantially older participants took far longer on the task; narrowing the age range reduces variance in baseline speed without targeting any experimental outcome, and the resulting samples are similar in age across conditions (Table~\ref{tab:exclusions}). Table~\ref{tab:prereg} maps each experiment to its pre-registration(s), pre-registered target sample size, and AsPredicted link.

\begin{table}[htbp]
\renewcommand{\arraystretch}{1.2}
\centering
\caption{Mapping of Experiments to Pre-Registrations}
\label{tab:prereg}
\footnotesize
\begin{tabular}{@{}lp{4.1cm}p{2.4cm}p{4.0cm}@{}}
\toprule
Experiment & Pre-registered content & Target sample & AsPredicted link \\
\midrule
Experiment 1a (\textit{Endog}) & Design, exclusion criteria, and the workload and cherry-picking hypotheses (\hypref{H1}, \hypref{H2}) & 100 & \url{https://aspredicted.org/ik5kp8.pdf} \\
Experiment 1b (\textit{Pre-commit}) & Rule pre-commitment; sequencing preferences and autonomy (\hypref{H3}, \hypref{H4}) & 100 & \url{https://aspredicted.org/y9qi9n.pdf} \\
Experiment 2a (three-arm) & Main treatment effects (\hypref{H5}); \newline heterogeneity and personality analyses & 300 total (100 per treatment) & \url{https://aspredicted.org/u88wg3.pdf} \newline \url{https://aspredicted.org/cu3wm6.pdf} \\
Experiment 2b (\textit{Invisible-FIFO}) & Queue visibility (\hypref{H6}) & 100 & \url{https://aspredicted.org/y9qi9n.pdf} \\
\bottomrule
\end{tabular}
\vspace{0.3em}

\begin{minipage}{0.92\textwidth}
\scriptsize \textit{Notes:} Sample sizes are pre-registered targets. Experiments~1b and~2b were pre-registered jointly in a single document.
\end{minipage}
\end{table}

Table~\ref{tab:exclusions} reports, for each condition, the number of workers who completed the experiment and the number excluded under each criterion: failure of the comprehension quiz (four or more errors, which no completing worker triggered in any wave), self-reported technical problems that genuinely interfered with task execution, and inattentiveness (an idle gap of more than 60 seconds between jobs, a screen added after pre-registration). The table also summarizes the demographic composition of each final sample.

\begin{table}[htbp]
\renewcommand{\arraystretch}{1.15}
\centering
\caption{Sample Construction: Completions and Exclusions by Condition}
\label{tab:exclusions}
\footnotesize
\begin{tabular}{llcccccccc}
\toprule
 & Condition & Completed & \multicolumn{2}{c}{Excluded} & Final sample & \multicolumn{4}{c}{Final-sample demographics} \\
\cmidrule(lr){4-5} \cmidrule(lr){7-10}
 & & & Quiz or technical & Inattentive &   & Female & Age & College+ & \$60k+ \\
\midrule
Exp.~1a & \textit{Endog} & 103 & 0 & 3 & 100 & 50\% & 33.2 & 55\% & 31\% \\
Exp.~1b & \textit{Pre-commit} & 118 & 6 & 1 & 111 & 50\% & 32.0 & 60\% & 44\% \\
Exp.~2a & \textit{Endog} & 107 & 3 & 3 & 101 & 50\% & 32.8 & 68\% & 39\% \\
Exp.~2a & \textit{Exog-FIFO} & 113 & 4 & 2 & 107 & 49\% & 32.3 & 64\% & 37\% \\
Exp.~2a & \textit{Exog-EF} & 102 & 0 & 1 & 101 & 46\% & 32.7 & 56\% & 31\% \\
Exp.~2b & \textit{Invisible-FIFO} & 108 & 3 & 5 & 100 & 52\% & 31.2 & 69\% & 40\% \\
\bottomrule
\end{tabular}
\vspace{0.3em}

\begin{minipage}{0.96\textwidth}
\scriptsize \textit{Notes:} Quiz or technical = four or more comprehension-quiz errors, or a self-reported technical problem that interfered with the task (reports of minor lag are not excluded). Inattentive = at least one idle gap over 60 seconds between consecutive jobs in the main rounds. Demographics describe the final sample: College+ is an undergraduate degree or higher and \$60k+ is annual pre-tax income of \$60,000 or more, both among respondents who disclosed. Gender, education, and income do not differ across conditions (rank sum tests on all pairwise comparisons, $p>0.05$); the \textit{Invisible-FIFO} sample is about two years younger than the \textit{Endog} conditions (rank sum $p<0.05$). 
\end{minipage}
\end{table}

\section{Pilot Experiments}
\label{ec:pilot}
\FloatBarrier

Prior to running the main experiment, we conducted a pilot study ($n=56$) on Prolific in January 2026. The pilot used the same task and interface as the main experiment, but did not include the mistake-tolerance manipulation or the exogenous arrival-rate variation. Instead, all jobs arrived according to a fixed schedule across four main rounds (Rounds 1--4), with a practice round (Round~0) and no final round. We describe how the pilot affected the design of subsequent experiments below.\footnote{Additionally, before running Experiment~2a, we ran two further pilots (\url{https://aspredicted.org/kk6mg5.pdf}) in which workers were assigned to an imposed FIFO or an imposed easiest-first sequence. The pilots established that enforced sequencing worked as intended and suggested that treatment effects vary with baseline ability, which motivated pre-registering the ability-moderation analysis of Experiment~2a.}

\subsubsection*{Completion times.}
The pilot established baseline completion times for each job type. Table~\ref{tab:pilot:times} reports the distribution of per-job completion times across Rounds~1--4. As expected, completion times increase with job size: median times were approximately 7, 14, and 24 seconds for Small (1~item), Medium (3~items), and Large (10~items) jobs, respectively. These completion times informed the timing parameters (job arrival schedule) of the main experiment.

\begin{table}[htbp]
\renewcommand{\arraystretch}{1.1}
\caption{Pilot: Completion Time (sec) by Job Type}
\label{tab:pilot:times}
\footnotesize
\centering
\begin{tabular}{lcccccc}
\toprule
Job Type & $N$ & Mean & Median & P75 & P90 & P95 \\
\midrule
Small (1 item) & 730 & 9.7 & 7.2 & 9.8 & 14.6 & 19.4 \\
Medium (3 items) & 727 & 23.3 & 13.8 & 20.5 & 42.7 & 66.7 \\
Large (10 items) & 779 & 37.8 & 24.0 & 37.6 & 67.2 & 102.5 \\
\bottomrule
\end{tabular}
\end{table}

\subsubsection*{Calibrating arrival times.}
A key design goal for the main experiment was to ensure that participants always had a meaningful choice of which job to work on next. If jobs arrive too slowly, the queue empties and no selection decision exists; if jobs arrive too quickly, the queue is always long and arrival-rate variation becomes irrelevant. We used the pilot completion-time distributions to calibrate the inter-arrival times in the main experiment. Specifically, we set inter-arrival times so that new jobs would arrive while even the fastest workers (those at or below the 25th percentile of completion times) were still working on their current job. In the main experiment, the 9-job (low arrival rate) rounds space arrivals approximately every 10 seconds over the first 80 seconds, while the 12-job (high arrival rate) rounds space them every 7 seconds over the same window. This ensures that at any given moment, most participants face a queue with multiple jobs of varying difficulty from which to choose.

\subsubsection*{Cherry-picking behavior.}
The pilot confirmed that participants do deviate from FIFO order when given the opportunity. On average, participants committed 2.2 FIFO violations per round, and approximately 49\% of participants violated FIFO at least once in any given round. Violations declined across rounds (from 2.6 in Round~1 to 1.8 in Round~4), suggesting that cherry-picking decreases with experience. These patterns motivated our focus on cherry-picking as a primary behavioral outcome.

\subsubsection*{Design refinements.}
Based on the pilot, we made several refinements for the main experiment: (i)~we added the mistake-tolerance manipulation to examine how error consequences affect job selection; (ii)~we introduced two arrival-rate conditions (9 vs.\ 12 jobs) to create exogenous variation in workload; (iii)~we added a final round with fixed job order to provide a post-treatment baseline measure of ability; and (iv)~we included a non-binding reference timer to create additional variation in perceived time pressure without imposing a hard deadline.

\section{Experimental Instructions}
\label{ec:instructions}
\FloatBarrier

This section reproduces the instructions and key screens shown to participants during the experiment. The experiment was programmed in oTree and administered online via Prolific.

\subsection*{Instructions}
\label{ec:instructions:main}

After providing informed consent (see \S\ref{ec:instructions:consent}), participants saw the following instructions:

\begin{instructionbox}
In this task you will be asked to play the role of a service worker completing retail jobs.

Each job will include a list of items, for example, bread, milk and eggs. You will need to find these items in the list and add them to the cart. You will receive 5 cents for each completed job. This is in addition to the participation fee of \$4.

There are three types of jobs:
\begin{itemize}[nosep]
    \item Small job, includes 1 item
    \item Medium job, includes 3 items
    \item Large job, includes 10 items
\end{itemize}

Jobs will arrive throughout the round and will be added to the queue. Your task is to complete all the jobs (You cannot skip any jobs). However, you are free to choose the sequence in which the jobs are completed.

\textbf{Time:} There is no time limit for this task. A timer will be displayed for reference only -- most people are able to finish within the time shown. You are free to continue working even after the timer runs out.
\end{instructionbox}

After reading the instructions, participants answered comprehension questions. They could not proceed until all answers were correct.

\subsection*{Practice Round}
\label{ec:instructions:practice}

Before the main task, participants completed a practice round:

\begin{instructionbox}
\textbf{Practice Round}

This is a practice round to help you get familiar with the task. You will complete 5 jobs of varying size in a pre-determined order.

You will receive \textbf{5 cents per correctly completed job}. If you make any mistakes on a job, you will need to redo it.
\end{instructionbox}

In the practice round, all five jobs were available from the start and had to be completed in a fixed (sequential) order. This familiarized participants with the shopping interface without introducing selection decisions.

\subsection*{Main Rounds (Rounds 1--4)}
\label{ec:instructions:main_rounds}

At the start of each main round (Rounds 1--4), participants saw the following message:

\begin{instructionbox}
You can now continue with the next round. As before, you will complete jobs for payment (5 cents per correctly completed job).
\end{instructionbox}

In the main rounds, unlike the practice round, participants were free to select which job to work on next from the available queue.

\subsection*{Mistake-Tolerance Manipulation}
\label{ec:instructions:tolerance}

Before the two rounds with mistake tolerance, participants saw the following screen (with a comprehension check):

\begin{instructionbox}
\textbf{Mistake Tolerance}

\textbf{For the next two rounds}, the following rules apply:

\begin{itemize}[nosep]
    \item For \textbf{Medium and Large jobs only}, you can make \textbf{up to 1 mistake} and still receive payment (5 cents).
    \item A ``mistake'' means one item error -- for example: missing 1 item, having 1 wrong quantity, or substituting 1 wrong item for a required item.
    \item If you make \textbf{more than 1 mistake} on a Medium or Large job, you will need to redo it.
    \item \textbf{Small jobs (1 item) must be completed perfectly} -- no mistakes allowed.
\end{itemize}
\end{instructionbox}

Before the two rounds \emph{without} mistake tolerance, participants saw:

\begin{instructionbox}
\textbf{No Mistake Tolerance}

\textbf{For the next two rounds}, the following rules apply:

\begin{itemize}[nosep]
    \item You must \textbf{complete each job perfectly} (without any mistakes) to receive payment (5 cents per job).
    \item If you make \textbf{any mistakes} on a job, you will need to redo it.
\end{itemize}
\end{instructionbox}

\subsection*{Final Round (Round 5)}
\label{ec:instructions:final}

Before the final round, participants saw:

\begin{instructionbox}
You can now continue with the next round. \textbf{In this final round, you will not be able to reorder jobs. That is, you will be required to complete the jobs in a pre-determined order.} As before, you will complete jobs for payment (5 cents per correctly completed job).

\textbf{No Mistake Tolerance:} In this final round, you must complete each job perfectly (without any mistakes) to receive payment. If you make any mistake, you will need to redo the job.
\end{instructionbox}

Like the practice round, the final round presented five jobs in a fixed order, providing a post-treatment measure of baseline ability.

\subsection*{Task Interface}
\label{ec:instructions:interface}

\begin{figure}[htbp]
\centering
\caption{Screenshot: Job List}
\includegraphics[width=0.75\textwidth]{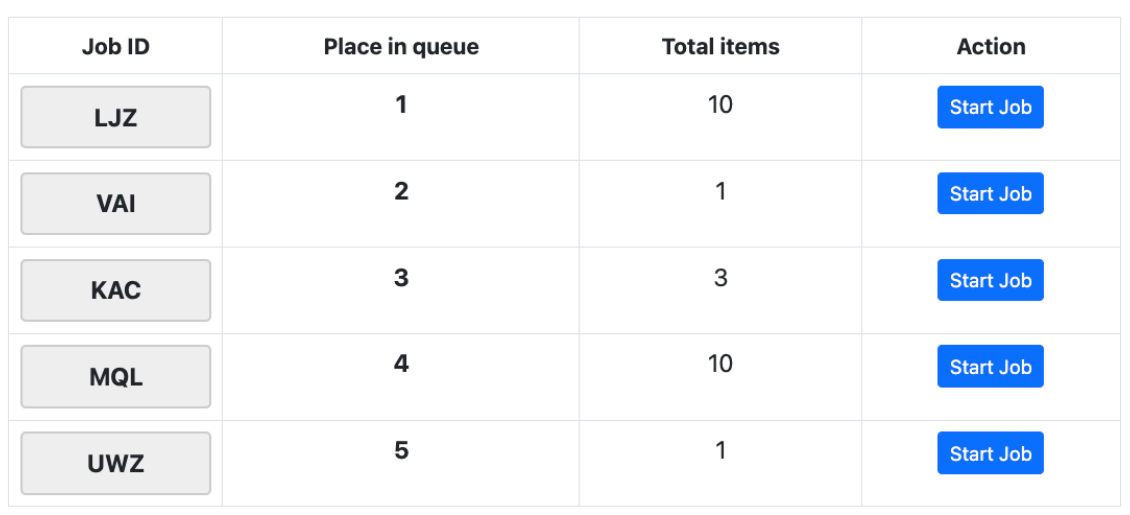}
\label{fig:ec:joblist}
\end{figure}

\paragraph{Job List.} The main task screen displayed a table of available jobs, showing the Job~ID (a random 3-letter code), place in the queue, total number of items, and a ``Start Job'' button. In the main rounds (Rounds 1--4), participants could click any available job; in the practice and final rounds, only the next job in the sequence was clickable. Figure~\ref{fig:ec:joblist} shows a screenshot of the job list.

\begin{figure}[t]
\centering
\caption{Screenshot: Shopping Screen}
\includegraphics[width=0.75\textwidth]{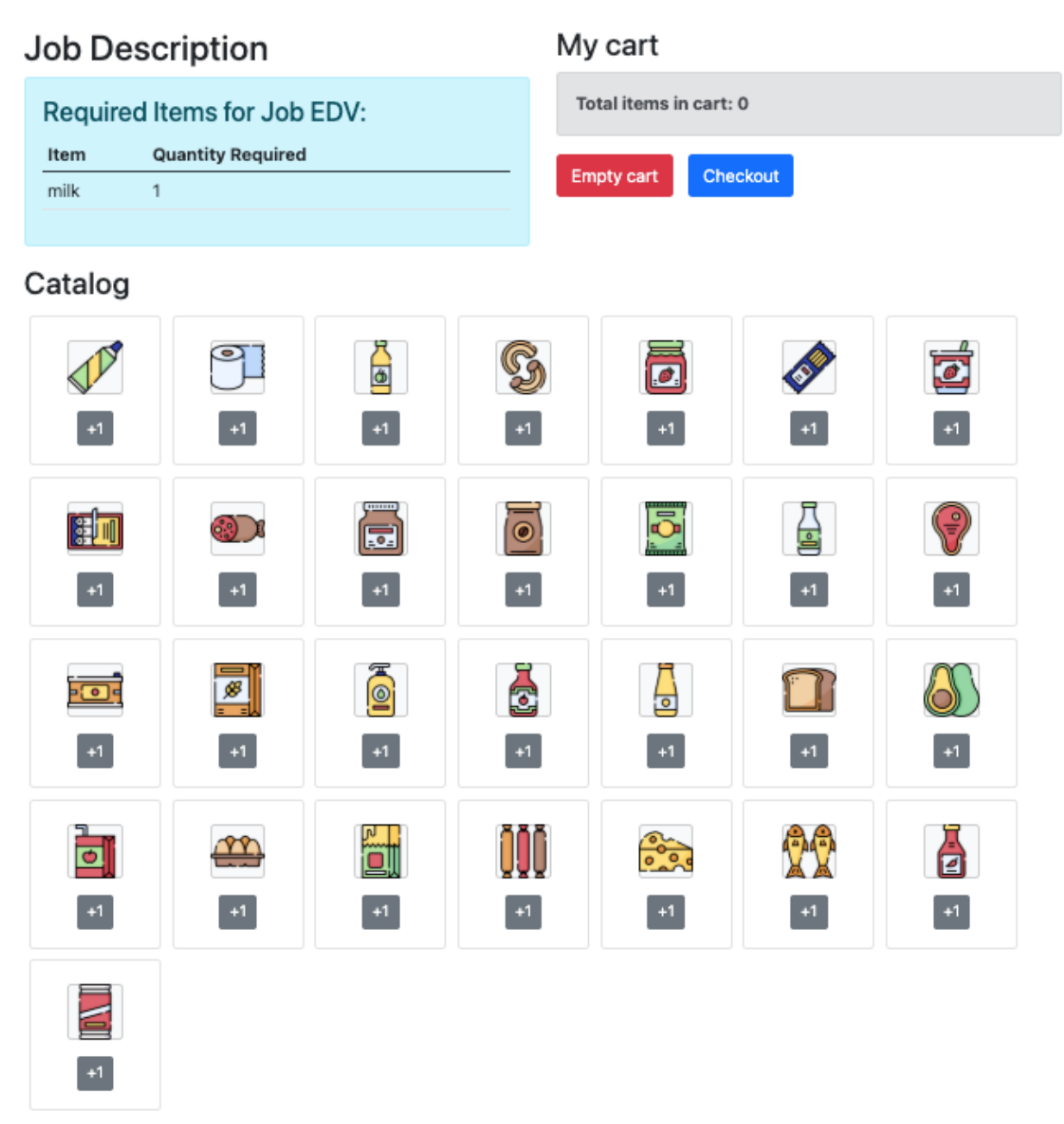}
\label{fig:ec:cart}
\end{figure}

\paragraph{Shopping Screen.} Upon selecting a job, participants saw (i)~a Job Description panel listing the required items and quantities, (ii)~a My Cart panel showing the current cart contents with ``Empty cart'' and ``Checkout'' buttons, and (iii)~a Catalog of product icons, each with a ``+1'' button to add the item to the cart. Figure~\ref{fig:ec:cart} shows a screenshot of the shopping interface.

\paragraph{Checkout Feedback.} After clicking ``Checkout,'' participants saw a comparison of required vs.\ collected items. If all items were correct, a ``Return to Job List'' button appeared. If errors were found, participants saw one of the following messages:

In zero-tolerance rounds (or if more than one mistake was made in error-tolerant rounds):
\begin{instructionbox}
[Table comparing required vs.\ collected items, with errors highlighted]

\textbf{Redo Job}
\end{instructionbox}

In tolerance rounds, if exactly 1 mistake was made on a Medium or Large job:
\begin{instructionbox}
[Table comparing required vs.\ collected items, with error highlighted]

\textbf{Note:} You made 1 mistake on this job, but you can still continue and receive payment (5 cents) for this job.

\textbf{Continue}
\end{instructionbox}

\subsection*{Manipulation Check}
\label{ec:instructions:manipulation}

After each of the four main rounds, participants answered two manipulation-check questions on a seven-point scale (1 = ``Not at all'' to 7 = ``Extremely''):

\begin{instructionbox}
Please think about the task you just completed and rate your agreement with the following statements:
\begin{enumerate}[nosep]
    \item To what extent did you feel pressed for time?
    \item To what extent are you feeling fatigued?
\end{enumerate}
\end{instructionbox}

\subsection*{Round Summary}
\label{ec:instructions:summary}

After each round, participants saw a summary screen displaying the number of jobs completed (out of total), the bonus earned in that round, and the cumulative bonus. After the final round, the screen additionally showed the total payment (participation fee plus bonus).

\subsection*{Sequencing-Rule Choice (Experiment 1b)}
\label{ec:instructions:precommit}

\begin{figure}[htbp]
\centering
\caption{Screenshot: Sequencing-Rule Choice (Experiment 1b)}
\label{fig:precommit_choice}
\includegraphics[width=0.75\textwidth]{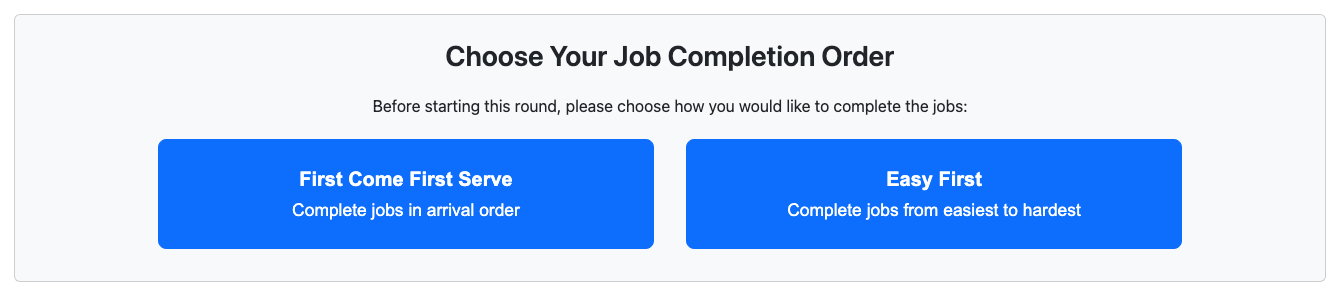}
\end{figure}

In the \textit{Pre-commit} condition (Experiment~1b), the task instructions and interface were identical to Experiment~1a, except that before each main round participants chose a sequencing rule on the screen shown in Figure~\ref{fig:precommit_choice}. \textit{First Come First Serve} (``complete jobs in arrival order'') corresponds to FIFO in the main text, and \textit{Easy First} (``complete jobs from easiest to hardest'') corresponds to EF. The chosen rule was enforced for the remainder of the round: only the job consistent with the rule could be selected from the queue. A fresh choice was made before each round.

\subsection*{Personality Measures (Experiment 2a)}
\label{ec:instruments}

In Experiment~2a, the following three instruments were administered after the consent page and before the task instructions; in Experiment~1 the same battery was administered at the end of the session. The order of items within the BFI-10 and within the NFCS-15 was randomized for each participant.

\subsubsection*{BFI-10 \citep{rammstedt2007}.} Participants rated how well each of ten statements describes them, completing the stem ``I see myself as someone who\ldots'', on a five-point scale (Disagree strongly, Disagree a little, Neither agree nor disagree, Agree a little, Agree strongly). The ten statements were: \emph{is reserved}; \emph{is generally trusting}; \emph{tends to be lazy}; \emph{is relaxed, handles stress well}; \emph{has few artistic interests}; \emph{is outgoing, sociable}; \emph{tends to find fault with others}; \emph{does a thorough job}; \emph{gets nervous easily}; \emph{has an active imagination}. Each Big Five dimension is measured by two items, one of which is reverse-coded.

\subsubsection*{Need for Closure \citep[NFCS-15;][]{roets2011}.} Participants rated their agreement with 15 statements on a six-point scale (1 = Completely disagree, \ldots, 6 = Completely agree):
\begin{enumerate}[nosep]
\item I don't like situations that are uncertain.
\item I dislike questions which could be answered in many different ways.
\item I find that a well ordered life with regular hours suits my temperament.
\item I feel uncomfortable when I don't understand the reason why an event occurred in my life.
\item I feel irritated when one person disagrees with what everyone else in a group believes.
\item I don't like to go into a situation without knowing what I can expect from it.
\item When I have made a decision, I feel relieved.
\item When I am confronted with a problem, I'm dying to reach a solution very quickly.
\item I would quickly become impatient and irritated if I would not find a solution to a problem immediately.
\item I don't like to be with people who are capable of unexpected actions.
\item I dislike it when a person's statement could mean many different things.
\item I find that establishing a consistent routine enables me to enjoy life more.
\item I enjoy having a clear and structured mode of life.
\item I do not usually consult many different opinions before forming my own view.
\item I dislike unpredictable situations.
\end{enumerate}
The decisiveness subscale averages items 7--9, and the preference-for-structure subscale averages items 3, 12, and 13 (numbering as listed above; the presentation order was randomized).

\subsubsection*{Risk appetite.} A single question following \citet{dohmen2011}: ``How do you see yourself: are you a person who is generally willing to take risks, or do you try to avoid taking risks?'', answered on an eleven-point scale from 0 (``Completely unwilling to take risks'') to 10 (``Very willing to take risks'').

\subsection*{Informed Consent}
\label{ec:instructions:consent}

Before beginning the experiment, participants provided informed consent (HIRB Application No.: HIRB00018806). The consent form informed participants that (i)~the study examines how individuals make decisions in the context of service provision; (ii)~the study involves completing an individual task lasting approximately 20 minutes; (iii)~the study poses no more than minimal risk; (iv)~compensation consists of a fixed payment of \$4 plus a performance-based bonus; and (v)~participation is voluntary and can be discontinued at any time. The full consent form is available from the authors upon request.

\section{Additional Analyses and Robustness}
\label{ec:robustness}
\FloatBarrier

This section collects supplementary analyses that support the results in the main text. The relevant information for each robustness check is summarized below.

\medskip

\begin{center}
\scriptsize
\setlength{\tabcolsep}{4pt}
\begin{tabular}{L{1.5cm}L{1.4cm}lll}
\toprule
 EC Subsection & Relevant Section in Main Text & Hypothesis & Analysis & Tables/Figures \\
\midrule
\S\ref{ec:rob:full} & \S\ref{sec:exp1a:recruit} & \hypref{H1}, \hypref{H2} & Full sample analysis, without the attentiveness screen & Tables~\ref{tab:reg:H1:fullsample}, \ref{tab:reg:H2:fullsample} \\
\S\ref{ec:rob:est} & \S\ref{sec:exp1a:tests} & \hypref{H1}, \hypref{H2} & Within-worker estimates with subject fixed effects & Tables~\ref{tab:reg:H1:fe}, \ref{tab:reg:lpm} \\
\S\ref{ec:rob:controls} & \S\ref{sec:exp1a:tests} & \hypref{H1}, \hypref{H2} & Job-content controls & Tables~\ref{tab:reg:H1:sku}, \ref{tab:reg:H2:sku} \\
\S\ref{ec:rob:direction} & \S\ref{sec:exp1a:tests} & \hypref{H2} & FIFO Deviations decomposed by direction & Table~\ref{tab:reg:H2:direction} \\
\S\ref{ec:rob:speed} & \S\ref{sec:exp1a:addl} & \hypref{H2} & Cherry-picking and speed in error-tolerant rounds & Table~\ref{tab:reg:H2:time} \\
\S\ref{ec:rob:arrival} & \S\ref{sec:exp1a:addl} & \hypref{H1}, \hypref{H2} & Arrival-timing controls & Table~\ref{tab:reg:H2:arrival} \\
\S\ref{ec:mixers} & \S\ref{sec:exp1b:tests} & \hypref{H3} & Sequencing rule consistency in Experiment~1b & Figure~\ref{fig:ef_groups} \\
\S\ref{ec:rob:corr} & \S\ref{sec:exp2a:pers} & -- & Correlations among demographic and personality measures and outcomes & Tables~\ref{tab:cormatrix}, \ref{tab:cormatrix:endog} \\
\bottomrule
\end{tabular}
\end{center}

\smallskip

\subsection{Full-Sample Results}
\label{ec:rob:full}

This subsection supports the Experiment~1a analyses of \S\ref{sec:exp1a} (Tables~\ref{tab:reg:H1} and~\ref{tab:reg:H2}). The main-text regressions apply a post-hoc attentiveness filter that drops 3 of the 103 subjects in the raw dataset (see \S\ref{sec:exp1a:recruit} for details), leaving an analysis sample of $n=100$. Tables~\ref{tab:reg:H1:fullsample} and~\ref{tab:reg:H2:fullsample} reproduce Tables~\ref{tab:reg:H1} and~\ref{tab:reg:H2} on the full 103-subject sample, with no exclusions other than the pre-registered ones (failure of the comprehension quiz and self-reported technical issues). Adding back the 3 filtered subjects does not change any of the reported results.

\begin{table}[htbp]
\renewcommand{\arraystretch}{1.1}
\caption{Effect of Workload Salience on Performance and Cherry-Picking (Full Sample)}
\label{tab:reg:H1:fullsample}
\footnotesize
\centering
\begin{tabular}{lC{1.6cm}C{1.6cm}C{1.6cm}C{1.6cm}C{1.6cm}C{1.6cm}}
\toprule
& (1) & (2) & (3) & (4) & (5) & (6) \\
\multicolumn{1}{r}{\textbf{Dependent Variable:}} & \multicolumn{2}{c}{\emph{Job Completion Time (sec)}} & \multicolumn{2}{c}{\emph{Job Error (0/1)}} & \multicolumn{2}{c}{\emph{Cherry-Picked (0/1)}} \\
\cmidrule(lr){2-3} \cmidrule(lr){4-5} \cmidrule(lr){6-7}
\multicolumn{1}{r}{\textbf{Sample:}} & \multicolumn{2}{c}{All rounds} & \multicolumn{2}{c}{Error-tolerant rounds} & \multicolumn{2}{c}{All rounds} \\
\cmidrule(lr){2-3} \cmidrule(lr){4-5} \cmidrule(lr){6-7}
\emph{Log(Time Since Last Job Arrival + 1)} & 1.418*** &  & $-$0.010 &  & $-$0.253*** &  \\
 & (0.242) &  & (0.054) &  & (0.039) &  \\
\addlinespace
\emph{Job Arrival: $\leq$10 sec} &  & $-$5.034*** &  & 0.096 &  & 1.316*** \\
 &  & (0.920) &  & (0.213) &  & (0.208) \\
\addlinespace
\emph{Job Arrival: 10--30 sec} &  & $-$4.614*** &  & 0.050 &  & 1.354*** \\
 &  & (1.258) &  & (0.328) &  & (0.224) \\
\addlinespace
\emph{Job Arrival: 30--60 sec} &  & $-$1.890 &  & 0.364 &  & 0.825*** \\
 &  & (1.674) &  & (0.222) &  & (0.245) \\
\midrule
Estimator & OLS & OLS & Logit & Logit & Logit & Logit \\
Round FE & Yes & Yes & Yes & Yes & Yes & Yes \\
Job Type Controls & Yes & Yes & Yes & Yes & -- & -- \\
Baseline Ability Controls & Yes & Yes & Yes & Yes & Yes & Yes \\
Queue Length Control & Yes & Yes & Yes & Yes & Yes & Yes \\
Observations & 4,326 & 4,326 & 1,438 & 1,438 & 3,482 & 3,482 \\
Subjects & 103 & 103 & 103 & 103 & 103 & 103 \\
Adj.\ $R^{2}$ & 0.164 & 0.164 & -- & -- & -- & -- \\
\bottomrule
\end{tabular}
\vspace{0.3em}

\begin{minipage}{0.97\textwidth}
\setlength{\baselineskip}{0.1\baselineskip}
\scriptsize \textit{Notes:} Full sample, without the post-hoc attentiveness filter. Same specification as Table~\ref{tab:reg:H1} in the main text. Tests of directional hypotheses (\hypref{H1a} and \hypref{H1b}) use one-sided $p$-values; remaining comparisons use two-sided $p$-values. *** $p<0.01$, ** $p<0.05$, * $p<0.1$.
\end{minipage}
\end{table}

\begin{table}[htbp]
\renewcommand{\arraystretch}{1.1}
\caption{Cherry-Picking and Performance (Full Sample)}
\label{tab:reg:H2:fullsample}
\footnotesize
\centering
\begin{tabular}{lC{1.5cm}C{1.5cm}C{1.5cm}C{1.5cm}C{1.5cm}C{1.5cm}}
\toprule
\multicolumn{1}{r}{\textbf{Dependent Variable:}} & \multicolumn{3}{c}{\emph{Job Completion Time (sec)}} & \multicolumn{3}{c}{\emph{Job Error (0/1)}} \\
\cmidrule(lr){2-4} \cmidrule(lr){5-7}
\multicolumn{1}{r}{\textbf{Sample:}} & \multicolumn{3}{c}{Zero-tolerance rounds} & \multicolumn{3}{c}{Error-tolerant rounds} \\
\cmidrule(lr){2-4} \cmidrule(lr){5-7}
& (1) & (2) & (3) & (4) & (5) & (6) \\
\midrule
\emph{Cherry-Picked} & 0.709 & 0.631 & 0.123 & 0.565** & 0.562** & 0.233 \\
 & (1.525) & (1.430) & (2.647) & (0.275) & (0.284) & (0.561) \\
\midrule
Estimator & OLS & OLS & OLS & Logit & Logit & FE Logit \\
Round FE & Yes & Yes & Yes & Yes & Yes & Yes \\
Job Type Controls & Yes & Yes & Yes & Yes & Yes & Yes \\
Baseline Ability Controls & No & Yes & No & No & Yes & No \\
Subject FE & No & No & Yes & No & No & Yes \\
Observations & 1,734 & 1,734 & 1,734 & 1,190 & 1,190 & 624 \\
Subjects & 103 & 103 & 103 & 103 & 103 & 55 \\
Adj.\ $R^{2}$ & 0.098 & 0.113 & 0.128 & -- & -- & -- \\
\bottomrule
\end{tabular}
\vspace{0.3em}

\begin{minipage}{0.85\textwidth}
\setlength{\baselineskip}{0.1\baselineskip}
\scriptsize \textit{Notes:} Full sample, without the post-hoc attentiveness filter. Same specification as Table~\ref{tab:reg:H2} in the main text. One-sided $p$-values are reported for the directional prediction (\hypref{H2}); the remaining comparisons use two-sided $p$-values. *** $p<0.01$, ** $p<0.05$, * $p<0.1$.
\end{minipage}
\end{table}

\subsection{Alternative Within-Worker Specifications}
\label{ec:rob:est}

This subsection supports the within-worker estimates of \S\ref{sec:exp1a}, namely cols.~(5)--(6) of Table~\ref{tab:reg:H1} and col.~(6) of Table~\ref{tab:reg:H2}.

Columns~(5)--(6) of Table~\ref{tab:reg:H1} control for worker heterogeneity through baseline ability. Table~\ref{tab:reg:H1:fe} instead re-estimates them with subject fixed effects, using two estimators. The conditional (fixed-effects) logit in cols.~(1)--(2) absorbs all worker heterogeneity but retains only the 76 workers whose cherry-picking varies across their own jobs; the linear probability models in cols.~(3)--(4) retain all 100 workers. Both reproduce the main-text conclusion: cherry-picking rises sharply following a recent arrival (log-odds $-0.128$ and $-1.3$ percentage points per log unit, both $p<0.05$), and the three arrival bins are ordered as in Table~\ref{tab:reg:H1}.

In Table~\ref{tab:reg:lpm} we turn to the effect of cherry-picking on job errors and re-estimate col.~(6) of Table~\ref{tab:reg:H2} as a linear probability model. We do this because the conditional logit in the main text is identified only from the 54 workers who make at least one error, which makes it both imprecise and specific to a selected subsample. In contrast, the linear probability model is able to retain all 100 workers. The conclusion is unchanged: the within-worker effect of cherry-picking on errors is $1.9$ percentage points and is not statistically significant ($p=0.67$).

\begin{table}[tbp]
\renewcommand{\arraystretch}{1.1}
\caption{Workload Salience and Cherry-Picking: Subject Fixed Effects}
\label{tab:reg:H1:fe}
\footnotesize
\centering
\begin{tabular}{lC{1.9cm}C{1.9cm}C{1.9cm}C{1.9cm}}
\toprule
& (1) & (2) & (3) & (4) \\
\multicolumn{1}{r}{\textbf{Dependent Variable:}} & \multicolumn{4}{c}{\emph{Cherry-Picked (0/1)}} \\
\cmidrule(lr){2-5}
\multicolumn{1}{r}{\textbf{Estimator:}} & \multicolumn{2}{c}{FE Logit} & \multicolumn{2}{c}{Linear Probability} \\
\cmidrule(lr){2-3} \cmidrule(lr){4-5}
\emph{Log(Time Since Last Job Arrival + 1)} & $-$0.128** & & $-$0.013** & \\
 & (0.064) & & (0.006) & \\
\addlinespace
\emph{Job Arrival: $\leq$10 sec} & & 0.895*** & & 0.079*** \\
 & & (0.285) & & (0.024) \\
\addlinespace
\emph{Job Arrival: 10--30 sec} & & 1.100*** & & 0.108*** \\
 & & (0.293) & & (0.027) \\
\addlinespace
\emph{Job Arrival: 30--60 sec} & & 0.588** & & 0.051** \\
 & & (0.323) & & (0.026) \\
\midrule
Round FE & Yes & Yes & Yes & Yes \\
Subject FE & Yes & Yes & Yes & Yes \\
Queue Length Control & Yes & Yes & Yes & Yes \\
Observations & 2,483 & 2,483 & 3,379 & 3,379 \\
Subjects & 76 & 76 & 100 & 100 \\
Adj.\ $R^{2}$ & -- & -- & 0.387 & 0.390 \\
\bottomrule
\end{tabular}
\vspace{0.3em}

\begin{minipage}{0.8\textwidth}
\setlength{\baselineskip}{0.1\baselineskip}
\scriptsize \textit{Notes:} Same sample as cols.~(5)--(6) of Table~\ref{tab:reg:H1} (selections from mixed-difficulty queues), estimated with subject fixed effects in place of the baseline ability controls. Cols.~(1)--(2) report log-odds coefficients from a conditional (fixed-effects) logit, which retains only the subjects whose outcome varies across their own jobs; cols.~(3)--(4) report linear probability models, which retain all 100 workers. Standard errors clustered at the subject level in parentheses. The omitted job arrival category is $>60$ sec. One-sided $p$-values for the directional prediction (\hypref{H1b}); two-sided otherwise. *** $p<0.01$, ** $p<0.05$, * $p<0.1$.
\end{minipage}
\end{table}

\begin{table}[htbp]
\renewcommand{\arraystretch}{1.1}
\centering
\caption{Cherry-Picking and Job Errors: Linear Probability Model}
\label{tab:reg:lpm}
\footnotesize
\begin{tabular}{lC{2.6cm}}
\toprule
\multicolumn{1}{r}{\textbf{Dependent Variable:}} & \emph{Job Error (0/1)} \\
\cmidrule(lr){2-2}
\multicolumn{1}{r}{\textbf{Sample:}} & Error-tolerant rounds \\
\cmidrule(lr){2-2}
& (1) \\
\midrule
\emph{Cherry-Picked} & 0.019 \\
 & (0.043) \\
\midrule
Estimator & Linear Probability \\
Round FE & Yes \\
Job Type Controls & Yes \\
Subject FE & Yes \\
Observations & 1,156 \\
Subjects & 100 \\
Adj.\ $R^{2}$ & 0.231 \\
\bottomrule
\end{tabular}
\vspace{0.3em}

\begin{minipage}{0.35\textwidth}
\scriptsize \textit{Notes:} Same sample and specification as col.~(6) of Table~\ref{tab:reg:H2}, estimated as a linear probability model, which retains all 100 workers. Standard errors clustered at the subject level in parentheses. *** $p<0.01$, ** $p<0.05$, * $p<0.1$.
\end{minipage}
\end{table}

\subsection{Job-Content Controls}
\label{ec:rob:controls}

This subsection reports two robustness checks, which test whether the main-text results are driven by the particular items a job contained rather than by its difficulty class alone. Tables~\ref{tab:reg:H1:sku} and~\ref{tab:reg:H2:sku} rerun the main-text regressions with 29 binary SKU indicators, one per catalog item, marking whether that item appears in the job. These indicators replace the job-type controls, which are collinear SKU indicators: the sum of the indicators is equal to the number of distinct items in the job, and that count defines the job type. The job-content controls are a much finer description of the work than the three difficulty classes, since they distinguish between jobs of the same size that contain different items. For example, ``butter'' in our catalog list has significantly higher error rates and longer picking times than most other items. Nevertheless, both these analyses produce essentially the same results as our main specifications reported in the main text.

Table~\ref{tab:reg:H1:sku} reproduces the completion-time columns of Table~\ref{tab:reg:H1}. The speed result is unchanged: the coefficient on \emph{Log(Time Since Last Job Arrival + 1)} remains positive and significant at $0.883$ (col.~1, vs.\ $1.370$ in Table~\ref{tab:reg:H1}), and the two most recent arrival bins remain roughly three seconds faster than the omitted bin (col.~2).

\begin{table}[htbp]
\renewcommand{\arraystretch}{1.1}
\centering
\caption{Workload Salience and Job Completion Time, Controlling for Job Content}
\label{tab:reg:H1:sku}
\footnotesize
\begin{tabular}{lC{2.0cm}C{2.0cm}}
\toprule
\multicolumn{1}{r}{\textbf{Dependent Variable:}} & \multicolumn{2}{c}{\emph{Job Completion Time (sec)}} \\
\cmidrule(lr){2-3}
\multicolumn{1}{r}{\textbf{Sample:}} & \multicolumn{2}{c}{All rounds} \\
\cmidrule(lr){2-3}
& (1) & (2) \\
\midrule
\emph{Log(Time Since Last Job Arrival + 1)} & 0.883*** &  \\
 & (0.271) &  \\
\addlinespace
\emph{Last Job Arrival: $\leq$10 sec} &  & $-$2.765*** \\
 &  & (0.926) \\
\addlinespace
\emph{Last Job Arrival: 10--30 sec} &  & $-$2.911*** \\
 &  & (1.079) \\
\addlinespace
\emph{Last Job Arrival: 30--60 sec} &  & $-$0.845 \\
 &  & (1.539) \\
\midrule
Estimator & OLS & OLS \\
Round FE & Yes & Yes \\
Queue Length Controls & Yes & Yes \\
Baseline Ability Controls & Yes & Yes \\
SKU Indicators & Yes & Yes \\
Observations & 4,200 & 4,200 \\
Subjects & 100 & 100 \\
Adj.\ $R^{2}$ & 0.229 & 0.229 \\
\bottomrule
\end{tabular}
\vspace{0.3em}

\begin{minipage}{0.58\textwidth}
\setlength{\baselineskip}{0.1\baselineskip}
\scriptsize \textit{Notes:} Same sample, variables, and specifications as cols.~(1)--(2) of Table~\ref{tab:reg:H1}, with the job-type controls replaced by 29 binary SKU indicators (one per catalog item). Standard errors clustered at the subject level in parentheses. The omitted job arrival category is $>60$ sec. Tests of the directional prediction (\hypref{H1a}) use one-sided $p$-values. *** $p<0.01$, ** $p<0.05$, * $p<0.1$.
\end{minipage}
\end{table}

Table~\ref{tab:reg:H2:sku} leaves every conclusion of Table~\ref{tab:reg:H2} intact. Cherry-picking remains unrelated to completion time (cols.~1--3), remains associated with more errors between workers (log-odds $0.58$ in both cols.~4--5, both $p<0.05$), and remains indistinguishable from zero once subject fixed effects are included (col.~6). The selection interpretation of \hypref{Result 2} is thus not an artifact of job composition.

\begin{table}[htbp]
\renewcommand{\arraystretch}{1.1}
\centering
\caption{Cherry-Picking and Performance, Controlling for Job Content}
\label{tab:reg:H2:sku}
\footnotesize
\begin{tabular}{lC{1.5cm}C{1.5cm}C{1.5cm}C{1.5cm}C{1.5cm}C{1.5cm}}
\toprule
\multicolumn{1}{r}{\textbf{Dependent Variable:}} & \multicolumn{3}{c}{\emph{Job Completion Time (sec)}} & \multicolumn{3}{c}{\emph{Job Error (0/1)}} \\
\cmidrule(lr){2-4} \cmidrule(lr){5-7}
\multicolumn{1}{r}{\textbf{Sample:}} & \multicolumn{3}{c}{Zero-tolerance rounds} & \multicolumn{3}{c}{Error-tolerant rounds} \\
\cmidrule(lr){2-4} \cmidrule(lr){5-7}
& (1) & (2) & (3) & (4) & (5) & (6) \\
\midrule
\emph{Cherry-Picked} & $-$0.488 & $-$0.369 & $-$1.769 & 0.579** & 0.579** & $-$0.010 \\
 & (1.675) & (1.602) & (2.992) & (0.302) & (0.308) & (0.429) \\
\midrule
Estimator & OLS & OLS & OLS & Logit & Logit & FE Logit \\
Round FE & Yes & Yes & Yes & Yes & Yes & Yes \\
Subject FE & No & No & Yes & No & No & Yes \\
Baseline Ability Controls & No & Yes & No & No & Yes & No \\
SKU Indicators & Yes & Yes & Yes & Yes & Yes & Yes \\
Observations & 1,682 & 1,682 & 1,682 & 1,156 & 1,156 & 609 \\
Subjects & 100 & 100 & 100 & 100 & 100 & 54 \\
Adj.\ $R^{2}$ & 0.179 & 0.187 & 0.207 & -- & -- & -- \\
\bottomrule
\end{tabular}
\vspace{0.3em}

\begin{minipage}{0.86\textwidth}
\setlength{\baselineskip}{0.1\baselineskip}
\scriptsize \textit{Notes:} Same samples, variables, and specifications as Table~\ref{tab:reg:H2}, with the job-type controls replaced by 29 binary SKU indicators (one per catalog item) marking whether the item appears in the job. The job-type controls are dropped because they are collinear with the 29 indicators for job content. Standard errors clustered at the subject level in parentheses. One-sided $p$-values are reported for the pre-registered directional prediction (\hypref{H2}); the remaining comparisons use two-sided $p$-values. *** $p<0.01$, ** $p<0.05$, * $p<0.1$.
\end{minipage}
\end{table}

\subsection{FIFO Violations by Type}
\label{ec:rob:direction}

This subsection supports Table~\ref{tab:reg:H2} of \S\ref{sec:exp1a} by decomposing FIFO violations by type. A violation occurs whenever the worker selects a job that is not at the front of the queue; we split violations into \emph{Picked Easier} (a later job of lower complexity, i.e., cherry-picking), \emph{Same Difficulty}, and \emph{Picked Harder} (a later job of higher complexity, for example a Large job chosen when a Medium was at the front). Table~\ref{tab:reg:H2:direction} reruns the specifications of Table~\ref{tab:reg:H2} with these three dummies (omitted category: no violation), controlling for job content rather than job type so that jobs of the same size but different items are distinguished. The results are consistent with the main text. None of the three violation types is associated with completion time (cols.~1--3). For errors, only \emph{Picked Easier} is associated with more errors between workers (log-odds $0.56$, cols.~4--5, $p<0.05$), and it disappears within worker (col.~6). 

\begin{table}[htbp]
\renewcommand{\arraystretch}{1.1}
\caption{FIFO Violations and Performance, by Violation Direction}
\label{tab:reg:H2:direction}
\footnotesize
\centering
\begin{tabular}{lC{1.5cm}C{1.5cm}C{1.5cm}C{1.5cm}C{1.5cm}C{1.5cm}}
\toprule
\multicolumn{1}{r}{\textbf{Dependent Variable:}} & \multicolumn{3}{c}{\emph{Job Completion Time (sec)}} & \multicolumn{3}{c}{\emph{Job Error (0/1)}} \\
\cmidrule(lr){2-4} \cmidrule(lr){5-7}
\multicolumn{1}{r}{\textbf{Sample:}} & \multicolumn{3}{c}{Zero-tolerance rounds} & \multicolumn{3}{c}{Error-tolerant rounds} \\
\cmidrule(lr){2-4} \cmidrule(lr){5-7}
& (1) & (2) & (3) & (4) & (5) & (6) \\
\midrule
FIFO Violation: & & & & & & \\
\quad \emph{Picked Easier} & $-$0.182 & $-$0.057 & $-$1.545 & 0.560** & 0.555** & $-$0.074 \\
 & (1.750) & (1.677) & (3.548) & (0.324) & (0.329) & (0.463) \\
\addlinespace
\quad \emph{Same Difficulty} & $-$0.673 & $-$0.894 & $-$2.750 & $-$0.004 & $-$0.021 & 0.291 \\
 & (2.381) & (2.172) & (3.966) & (0.372) & (0.368) & (0.592) \\
\addlinespace
\quad \emph{Picked Harder} & 3.388 & 2.733 & 2.232 & 0.158 & 0.152 & 0.900 \\
 & (3.156) & (3.126) & (3.783) & (0.487) & (0.486) & (0.615) \\
\midrule
Estimator & OLS & OLS & OLS & Logit & Logit & FE Logit \\
Subject FE & No & No & Yes & No & No & Yes \\
Baseline Ability Controls & No & Yes & No & No & Yes & No \\
Round FE & Yes & Yes & Yes & Yes & Yes & Yes \\
Job Content Controls & Yes & Yes & Yes & Yes & Yes & Yes \\
Observations & 1,682 & 1,682 & 1,682 & 1,156 & 1,156 & 609 \\
Subjects & 100 & 100 & 100 & 100 & 100 & 54 \\
Adj.\ $R^{2}$ & 0.179 & 0.187 & 0.207 & -- & -- & -- \\
\bottomrule
\end{tabular}
\vspace{0.3em}

\begin{minipage}{0.87\textwidth}
\scriptsize \textit{Notes:} Direction decomposition of Table~\ref{tab:reg:H2} (omitted category: no violation). OLS coefficients in cols.~(1)--(3); logit log-odds in cols.~(4)--(6), with col.~(6) a fixed-effects (conditional) logit identified from the 54 workers with error variation (see \S\ref{ec:rob:est}), so its estimates are imprecise. Job content is controlled with 29 binary SKU indicators, one per catalog item, marking whether the item appears in the job. These replace the job-type controls used in Table~\ref{tab:reg:H2}, which cannot be included alongside them because the indicators sum to the number of distinct items in the job, and that count defines its type. Standard errors clustered at the subject level in parentheses. One-sided $p$-values for \emph{Picked Easier} (\hypref{H2}); two-sided otherwise. *** $p<0.01$, ** $p<0.05$, * $p<0.1$.
\end{minipage}
\end{table}

\subsection{Cherry-Picking and Completion Time}
\label{ec:rob:speed}

This subsection supports the additional analyses of \S\ref{sec:exp1a}: it tests whether cherry-picking buys time savings in the rounds where errors are tolerated, complementing the error results in cols.~(4)--(6) of Table~\ref{tab:reg:H2}.

To assess whether the higher error rate under cherry-picking reflects a speed-accuracy trade-off, Table~\ref{tab:reg:H2:time} re-estimates cols.~(4)--(6) of Table~\ref{tab:reg:H2} with completion time as the outcome on the identical sample (error-tolerant rounds, Medium and Large jobs, mixed-difficulty queues). Cherry-picking is associated with no change in completion time, so workers who cherry-pick make more errors without finishing any faster.

\begin{table}[htbp]
\renewcommand{\arraystretch}{1.1}
\caption{Cherry-Picking and Completion Time (Error-Tolerant Rounds)}
\label{tab:reg:H2:time}
\footnotesize
\centering
\begin{tabular}{lC{2.1cm}C{2.1cm}C{2.1cm}}
\toprule
\multicolumn{1}{r}{\textbf{Dependent Variable:}} & \multicolumn{3}{c}{\emph{Job Completion Time (sec)}} \\
\cmidrule(lr){2-4}
\multicolumn{1}{r}{\textbf{Sample:}} & \multicolumn{3}{c}{Error-tolerant rounds} \\
\cmidrule(lr){2-4}
& (1) & (2) & (3) \\
\midrule
\emph{Cherry-Picked} & $-$0.693 & $-$0.763 & $-$0.882 \\
 & (1.094) & (1.148) & (1.422) \\
\midrule
Estimator & OLS & OLS & OLS \\
Round FE & Yes & Yes & Yes \\
Job Type Controls & Yes & Yes & Yes \\
Baseline Ability Controls & No & Yes & No \\
Subject FE & No & No & Yes \\
Observations & 1,156 & 1,156 & 1,156 \\
Subjects & 100 & 100 & 100 \\
Adj.\ $R^{2}$ & 0.133 & 0.165 & 0.193 \\
\bottomrule
\end{tabular}
\vspace{0.3em}

\begin{minipage}{0.68\textwidth}
\setlength{\baselineskip}{0.1\baselineskip}
\scriptsize \textit{Notes:} Same specification as cols.~(1)--(3) of Table~\ref{tab:reg:H2}, estimated on the sample of cols.~(4)--(6): error-tolerant rounds, Medium and Large jobs, mixed-difficulty queues (OLS). Standard errors clustered at the subject level in parentheses. *** $p<0.01$, ** $p<0.05$, * $p<0.1$.
\end{minipage}
\end{table}

\subsection{Arrival-Timing Controls}
\label{ec:rob:arrival}

Table~\ref{tab:reg:H2:arrival} reproduces Table~\ref{tab:reg:H2}, adding \emph{Log(Time Since Last Job Arrival + 1)} as a control. The association between cherry-picking and errors is unchanged, and arrival timing affects completion times but not error rates. This means that the workload result (\hypref{Result 1}) and the selection mechanism (\hypref{Result 2}) operate independently, i.e., the selection of error-prone workers into EF sequences is not due to the increased urgency created by recent arrivals.

\begin{table}[htbp]
\renewcommand{\arraystretch}{1.1}
\caption{Cherry-Picking and Performance, Controlling for Arrival Timing}
\label{tab:reg:H2:arrival}
\footnotesize
\centering
\begin{tabular}{lC{1.5cm}C{1.5cm}C{1.5cm}C{1.5cm}C{1.5cm}C{1.5cm}}
\toprule
\multicolumn{1}{r}{\textbf{Dependent Variable:}} & \multicolumn{3}{c}{\emph{Job Completion Time (sec)}} & \multicolumn{3}{c}{\emph{Job Error (0/1)}} \\
\cmidrule(lr){2-4} \cmidrule(lr){5-7}
\multicolumn{1}{r}{\textbf{Sample:}} & \multicolumn{3}{c}{Zero-tolerance rounds} & \multicolumn{3}{c}{Error-tolerant rounds} \\
\cmidrule(lr){2-4} \cmidrule(lr){5-7}
& (1) & (2) & (3) & (4) & (5) & (6) \\
\midrule
\emph{Cherry-Picked} & 1.684 & 1.603 & $-$0.417 & 0.640** & 0.661** & 0.195 \\
 & (1.436) & (1.398) & (2.726) & (0.279) & (0.289) & (0.591) \\
\addlinespace
\emph{Log(Time Since Last Job Arrival + 1)} & 2.518*** & 2.162*** & 1.171** & 0.051 & 0.079 & 0.071 \\
 & (0.534) & (0.552) & (0.567) & (0.074) & (0.073) & (0.109) \\
\midrule
Estimator & OLS & OLS & OLS & Logit & Logit & FE Logit \\
Round FE & Yes & Yes & Yes & Yes & Yes & Yes \\
Job Type Controls & Yes & Yes & Yes & Yes & Yes & Yes \\
Baseline Ability Controls & No & Yes & No & No & Yes & No \\
Subject FE & No & No & Yes & No & No & Yes \\
Observations & 1,682 & 1,682 & 1,682 & 1,156 & 1,156 & 609 \\
Subjects & 100 & 100 & 100 & 100 & 100 & 54 \\
Adj.\ $R^{2}$ & 0.111 & 0.116 & 0.126 & -- & -- & -- \\
\bottomrule
\end{tabular}
\vspace{0.3em}

\begin{minipage}{0.92\textwidth}
\setlength{\baselineskip}{0.1\baselineskip}
\scriptsize \textit{Notes:} Same samples and specifications as Table~\ref{tab:reg:H2}, adding \emph{Log(Time Since Last Job Arrival + 1)} as a control. Standard errors clustered at the subject level in parentheses. One-sided $p$-values for the pre-registered directional prediction (\hypref{H2}); two-sided otherwise. *** $p<0.01$, ** $p<0.05$, * $p<0.1$.
\end{minipage}
\end{table}

\subsection{Rule Consistency in Experiment 1b}
\label{ec:mixers}

This subsection supports the Experiment~1b analyses of \S\ref{sec:exp1b} (\hypref{Result 3} and Figure~\ref{fig:precommit_rule}).

\begin{figure}[htbp]
\centering
\caption{Experiment 1b: Outcomes by Rule-Consistency Group}
\includegraphics[width=0.8\textwidth]{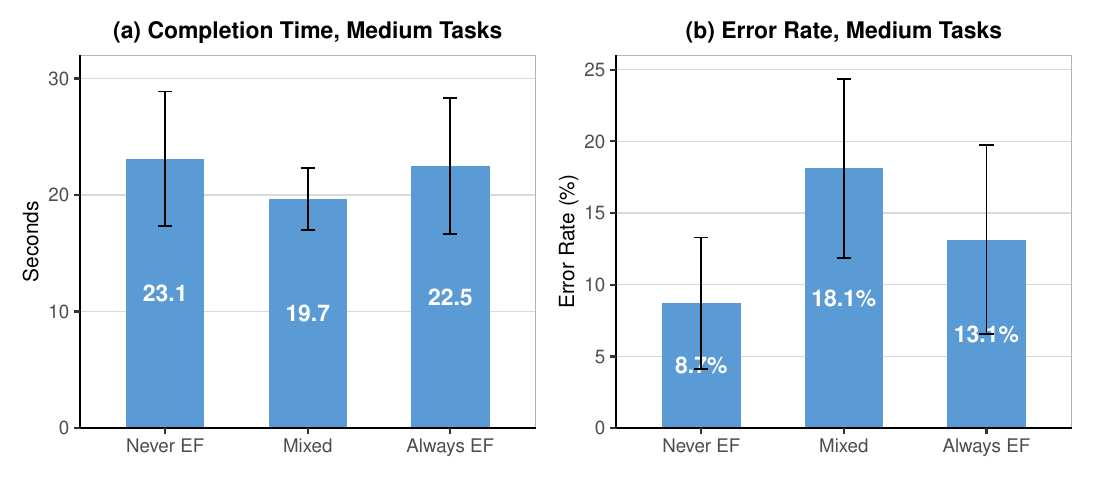}
\begin{minipage}{0.86\textwidth}\scriptsize
\textit{Notes:} \textit{Pre-commit} condition, Medium jobs. Panel (a): zero-tolerance rounds; panel (b): error-tolerant rounds. Never EF = chose FIFO in all four rounds ($n=38$ workers); Always EF = chose EF in all four rounds ($n=25$); Mixed = all others ($n=48$). Error bars are 95\% CIs, clustered at the subject level.
\end{minipage}\label{fig:ef_groups}
\end{figure}

The bimodal distribution of rule choices in Experiment~1b (\hypref{Result 3}) suggests a partition of the \textit{Pre-commit} sample into three groups: workers who never chose EF ($n=38$), workers who chose EF in every round ($n=25$), and workers who mixed the two rules across rounds ($n=48$). Figure~\ref{fig:ef_groups} compares the three groups on Medium jobs, using the same construction as Figure~\ref{fig:precommit_rule}.

Completion times are indistinguishable across the three groups ($23.1$, $19.7$, and $22.5$ seconds; Kruskal--Wallis test on subject-level means, $p=0.92$; all pairwise rank sum tests $p \geq 0.69$). Error rates are lowest among workers who never chose EF ($8.7\%$), highest among the mixers ($18.1\%$), and intermediate among consistent EF workers ($13.1\%$), but the differences are only marginally significant (Kruskal--Wallis, $p=0.16$; never-EF vs.\ mixers, rank sum test, $p=0.058$; the remaining pairwise comparisons $p \geq 0.34$). Pooling the two groups that choose EF at least once shows that workers who never chose EF make marginally fewer errors than the remaining workers (rank sum test, $p=0.077$). This analysis confirms that the selection mechanism, while present in the \textit{Pre-commit} treatment, is weaker than in the \textit{Endog} treatment. %

\subsection{Correlations among the Measures}
\label{ec:rob:corr}

This subsection supports the personality analyses of \S\ref{sec:exp2a} (Figure~\ref{fig:personality} and \hypref{Result 6}). Tables~\ref{tab:cormatrix} and~\ref{tab:cormatrix:endog} report the Pearson correlations among the demographic, personality, and behavioral measures used in \S\ref{sec:exp2a}, both across all three conditions and within the \textit{Endog} arm.

\begin{table}[htbp]
\centering
\caption{Correlations among the Measures and Behavioral Outcomes}
\label{tab:cormatrix}
\resizebox{\textwidth}{!}{%
\begin{tabular}{lrrrrrrrrrrrrrrrr}
\toprule
 & (1) & (2) & (3) & (4) & (5) & (6) & (7) & (8) & (9) & (10) & (11) & (12) & (13) & (14) & (15) & (16) \\
\midrule
(1) Female & -- \\
(2) Age & $-$.10* & -- \\
(3) Income & .04 & $-$.02 & -- \\
(4) Education & $-$.03 & .09 & $-$.02 & -- \\
(5) Risk appetite & $-$.13** & $-$.14** & $-$.08 & $-$.04 & -- \\
(6) Extraversion (Big 5) & .05 & $-$.08 & $-$.02 & $-$.00 & .33*** & -- \\
(7) Agreeableness (Big 5) & $-$.03 & .00 & $-$.04 & .03 & .12** & .33*** & -- \\
(8) Conscientiousness (Big 5) & .02 & .06 & $-$.08 & $-$.06 & $-$.01 & .21*** & .21*** & -- \\
(9) Neuroticism (Big 5) & .22*** & $-$.19*** & .11* & $-$.00 & $-$.24*** & $-$.35*** & $-$.25*** & $-$.39*** & -- \\
(10) Openness (Big 5) & .13** & $-$.11* & $-$.03 & $-$.00 & .05 & .07 & .05 & .17*** & $-$.09 & -- \\
(11) Need for closure & .04 & .04 & .02 & .03 & $-$.31*** & $-$.26*** & $-$.24*** & $-$.13** & .40*** & $-$.24*** & -- \\
(12) Pref.\ for structure & $-$.05 & $-$.01 & $-$.03 & .04 & $-$.33*** & $-$.12** & $-$.03 & .10* & .13** & $-$.19*** & .66*** & -- \\
(13) Decisiveness & .03 & $-$.11* & .08 & .02 & .02 & $-$.10* & $-$.18*** & $-$.11* & .26*** & $-$.10* & .63*** & .20*** & -- \\
(14) Cherry-picking rate & $-$.04 & $-$.01 & $-$.15 & $-$.09 & $-$.14 & $-$.01 & .03 & .14 & $-$.04 & $-$.08 & .01 & .21** & $-$.17* & -- \\
(15) Completion time & .00 & .06 & $-$.08 & $-$.02 & .11* & .10* & .05 & .08 & $-$.05 & .01 & .02 & .03 & .07 & .03 & -- \\
(16) Error rate & $-$.02 & .09 & $-$.01 & $-$.00 & .11* & .16*** & $-$.02 & .10* & $-$.11** & $-$.05 & .06 & .05 & .02 & $-$.06 & .09 & -- \\
\bottomrule
\end{tabular}}
\vspace{0.3em}

\begin{minipage}{\textwidth}
\scriptsize \textit{Notes:} Pearson correlations among the demographic and personality measures and the three behavioral outcomes (rows/columns 14--16), pooling all three conditions of Experiment~2a ($n=309$). Cherry-picking rate is defined only in the \textit{Endog} arm, so its cells use those $n=101$ workers. Completion time is the mean over zero-tolerance rounds; error rate is computed over Medium and Large jobs in error-tolerant rounds. Leading zeros omitted; lower triangle of the symmetric matrix. Big~5 traits from the BFI-10; Need for Closure and its subscales from the NFCS-15. *** $p<0.01$, ** $p<0.05$, * $p<0.1$.
\end{minipage}
\end{table}

\begin{table}[htbp]
\centering
\caption{Correlations among the Measures and Behavioral Outcomes (\textit{Endog} Arm)}
\label{tab:cormatrix:endog}
\resizebox{\textwidth}{!}{%
\begin{tabular}{lrrrrrrrrrrrrrrrr}
\toprule
 & (1) & (2) & (3) & (4) & (5) & (6) & (7) & (8) & (9) & (10) & (11) & (12) & (13) & (14) & (15) & (16) \\
\midrule
(1) Female & -- \\
(2) Age & $-$.16 & -- \\
(3) Income & $-$.08 & $-$.03 & -- \\
(4) Education & $-$.08 & .15 & $-$.04 & -- \\
(5) Risk appetite & $-$.19* & $-$.23** & $-$.10 & $-$.08 & -- \\
(6) Extraversion (Big 5) & .02 & $-$.04 & $-$.04 & $-$.02 & .34*** & -- \\
(7) Agreeableness (Big 5) & $-$.09 & .13 & $-$.08 & .08 & .08 & .41*** & -- \\
(8) Conscientiousness (Big 5) & .01 & .08 & $-$.17* & $-$.09 & $-$.13 & .22** & .30*** & -- \\
(9) Neuroticism (Big 5) & .22** & $-$.10 & .12 & $-$.02 & $-$.22** & $-$.47*** & $-$.49*** & $-$.35*** & -- \\
(10) Openness (Big 5) & .27*** & $-$.01 & $-$.12 & $-$.01 & .07 & .08 & .22** & .31*** & $-$.16 & -- \\
(11) Need for closure & .05 & .01 & .05 & .06 & $-$.31*** & $-$.28*** & $-$.32*** & $-$.15 & .38*** & $-$.17* & -- \\
(12) Pref.\ for structure & $-$.12 & $-$.07 & $-$.08 & .05 & $-$.34*** & $-$.06 & $-$.04 & .08 & .13 & $-$.15 & .65*** & -- \\
(13) Decisiveness & .09 & $-$.05 & .20** & .03 & .11 & $-$.13 & $-$.22** & $-$.19* & .12 & .02 & .54*** & $-$.01 & -- \\
(14) Cherry-picking rate & $-$.04 & $-$.01 & $-$.15 & $-$.09 & $-$.14 & $-$.01 & .03 & .14 & $-$.04 & $-$.08 & .01 & .21** & $-$.17* & -- \\
(15) Completion time & .08 & $-$.09 & $-$.03 & $-$.03 & .02 & .32*** & .13 & $-$.01 & $-$.13 & $-$.03 & .02 & .06 & .12 & .03 & -- \\
(16) Error rate & $-$.11 & $-$.03 & .01 & $-$.03 & .28*** & .30*** & $-$.04 & .02 & $-$.14 & $-$.06 & .01 & .05 & .03 & $-$.06 & .04 & -- \\
\bottomrule
\end{tabular}}
\vspace{-0.3em}

\begin{minipage}{\textwidth}
\scriptsize \textit{Notes:} As Table~\ref{tab:cormatrix}, but computed within the \textit{Endog} arm only ($n=101$), where all three outcomes (rows/columns 14--16) are observed for every worker. Leading zeros omitted; lower triangle of the symmetric matrix. *** $p<0.01$, ** $p<0.05$, * $p<0.1$.
\end{minipage}
\end{table}

\end{document}